\documentclass[]{aa}  

\usepackage{graphicx}
\usepackage{txfonts}

\usepackage[T1]{fontenc}
\usepackage{ae,aecompl}
\usepackage{subfig}
\usepackage{graphicx}	
\usepackage{amsmath}	
\usepackage{amssymb}	
\usepackage{amsbsy}
\usepackage{color}
\usepackage{xcolor}
\usepackage{CJKutf8}

\usepackage{booktabs}
\usepackage{lipsum}
\usepackage{rotating}
\usepackage{hyperref}
\usepackage{xspace}
\newcommand{\dr}{{\mathrm{d}}}
\newcommand{\planck}{{\it{Planck}}}

\newcommand{\euclid}{{\it{Euclid}}}
\newcommand{\bpe}{\left\langle b_y P_{\mathrm{e}}\right\rangle}
\newcommand{\bg}{\left\langle b_{\mathrm{g}}\right\rangle}
\newcommand{\cgy}{\left\langle c_{\mathrm{g}y}\right\rangle}
\newcommand{\cgk}{\left\langle c_{\mathrm{g}\kappa}\right\rangle}
\newcommand{\halofit}{\textsc{halofit}\xspace}
\newcommand{\healpix}{\textsc{healpix}\xspace}

\newcommand{\Mead}[1]{{\color{cyan}({Mead: #1})}}

\newcommand{\revised}[1]{{\color{black}{#1}}}
\newcommand{\revisednew}[1]{{\color{black}{#1}}}

\begin{document}

   \title{Probing galaxy bias and intergalactic gas pressure with KiDS Galaxies-tSZ-CMB lensing cross-correlations}

    \authorrunning{Yan et al.}

   \author{Ziang Yan\protect\begin{CJK*}{UTF8}{gbsn}    (颜子昂)\end{CJK*}
          \inst{1}\thanks{E-mail:yanza15@phas.ubc.ca}
            \and Ludovic van Waerbeke\inst{1}\thanks{E-mail:waerbeke@phas.ubc.ca}
            \and Tilman Tr\"{o}ster\inst{2}
            \and Angus H. Wright\inst{3}
            \and David Alonso\inst{4}
            \and Marika Asgari\inst{2}
            \and Maciej Bilicki\inst{5}
            \and Thomas Erben\inst{6}
            \and Shiming Gu \protect\begin{CJK*}{UTF8}{gbsn} (顾时铭)\end{CJK*}\inst{1}
            \and Catherine Heymans\inst{2}
            \and Hendrik Hildebrandt \inst{3}
            \and Gary Hinshaw \inst{1}
            \and Nick Koukoufilippas \inst{4}
            \and Arun Kannawadi \inst{7}
            \and Konrad Kuijken \inst{8}
            \and Alexander Mead \inst{2}
            \and HuanYuan Shan \inst{9}
          }

   \institute{Department of Physics and Astronomy, University of British Columbia, 6224 Agricultural Road, Vancouver, BC, V6T 1Z1, Canada\and
             Institute for Astronomy, University of Edinburgh, Royal Observatory, Blackford Hill, Edinburgh, EH9 3HJ, UK
             \and
             Ruhr University Bochum, Faculty of Physics and Astronomy, Astronomical Institute (AIRUB), German Centre for Cosmological Lensing, 44780 Bochum, Germany
             \and
             Department of Physics, University of Oxford, Denys Wilkinson Building, Keble Road, Oxford OX1 3RH, United Kingdom
             \and 
             Center for Theoretical Physics, Polish Academy of Sciences, al. Lotników 32/46, 02-668 Warsaw, Poland
             \and
             Argelander-Institut f\"ur Astronomie, Auf dem Hügel 71, 53121 Bonn, Germany
             \and 
             Department of Astrophysical Sciences, Princeton University, 4 Ivy Lane, Princeton, NJ 08544, USA
             \and 
             Leiden Observatory, Leiden University, P.O.Box 9513, 2300RA Leiden, The Netherlands
             \and 
             $^{1}$Shanghai Astronomical Observatory (SHAO), Nandan Road 80, Shanghai 200030, China $^{2}$University of Chinese Academy of Sciences, Beijing 100049, China
             }

   \date{}

 
  \abstract{We constrain the redshift dependence of gas pressure bias $\left\langle b_{y} P_{\mathrm{e}}\right\rangle$ (bias-weighted average electron pressure), which characterises the thermodynamics of intergalactic gas, through a combination of cross-correlations between galaxy positions and the thermal Sunyaev-Zeldovich (tSZ) effect, as well as galaxy positions and the gravitational lensing of the cosmic microwave background (CMB). The galaxy sample is from the fourth data release of the Kilo-Degree Survey (KiDS). The tSZ $y$ map and the CMB lensing map are from the {\textit{Planck}} 2015 and 2018 data releases, respectively. The measurements are performed in five redshift bins with $z\lesssim1$. With these measurements, combining galaxy-tSZ and galaxy-CMB lensing cross-correlations allows us to break the degeneracy between galaxy bias and gas pressure bias, and hence constrain them simultaneously. In all redshift bins, the best-fit values of $\bpe$ are at a level of $\sim 0.3\, \mathrm{meV/cm^3}$ and increase slightly with redshift. The galaxy bias is consistent with unity in all the redshift bins. Our results are not sensitive to the non-linear details of the cross-correlation, which are smoothed out by the {\textit{Planck}} beam. Our measurements are in agreement with previous measurements as well as with theoretical predictions. We also show that our conclusions are not changed when CMB lensing is replaced by galaxy lensing, which shows the consistency of the two lensing signals despite their radically different redshift ranges. This study demonstrates the feasibility of using CMB lensing to calibrate the galaxy distribution such that the galaxy distribution can be used as a mass proxy without relying on the precise knowledge of the matter distribution.}

     \titlerunning{KiDS-tSZ-CMB Lensing}
     
\maketitle
%
\section{Introduction}

The study of large-scale structure (LSS) is a major topic in modern cosmology. The theoretical framework of LSS in the linear regime is well established and has been constrained by multiple observations (see, for example, \citet{dodelson2020modern} for a detailed description). On small scales ($\sim 1$ Mpc), the growth of structure is driven by the combination of the non-linear gravitational collapse and baryonic processes in the intergalactic gas \citep{van2011effects, semboloni2011quantifying, fedelia2014clustering, mead2015accurate}. Although the latter is challenging to model, an increasing number of multi-wavelength sky surveys reach high redshifts and high angular resolution \citep[for example]{catinella2010galex, heymans2012cfhtlens,de2013kilo,abbott2016dark,2018PASJ...70S...4A}, which extend our understanding of the late-time history of the Universe and make us sensitive to subtle and complicated small-scale physics. In addition, observations of different `tracers' make it possible to study different aspects of LSS. In summary, it is a golden age dominated by surveys that shed light on the role of small-scale physics in LSS formation and evolution.

For many years, cross-correlations between different LSS tracers have been an important tool for helping us understand relations between underlying physics \citep{hill2014detection, van2014detection, kirk2016cross, hojjati2017cross, singh2017cross, ammazzalorso2020detection}. 
Compared to other LSS tracers, the galaxy distribution is easier to measure with high precision. Cross-correlations between galaxy positions and LSS has proved to be a powerful tool for studying different properties of LSS. For example, the cross-correlation between galaxy positions and cosmic infrared background (CIB) has been used to probe the properties of dust in star-forming galaxies \citep{Serra_2014}; the cross-correlation between galaxy position and 21 cm emission is useful for studying the cosmic reionisation history \citep{Lidz_2008}; \citet{Kuntz_2015} probes the cross-correlation between galaxy positions and cosmic microwave background (CMB) lensing to study the galaxy bias and lensing amplitude. In this study, we focus on the cross-correlation between the galaxy distribution and the thermal Sunyaev-Zeldovich (tSZ) effect (denoted as the `$gy$' cross-correlation hereafter) to probe the `gas pressure bias', defined as the multiplication of the mean electron pressure and gas bias, the ratio of the gas overdensity to the mass overdensity, as a proxy of intergalactic gas properties.

The tSZ effect \citep{zeldovich1969interaction, sunyaev1972observations} is the distortion of the CMB energy spectrum due to inverse Compton scattering by high-energy electrons; tSZ effect is therefore a tracer of the projected intergalactic gas pressure. Since warm to hot intergalactic gas is typically concentrated in galaxy clusters, one can use the tSZ effect to detect galaxy clusters \citep{ade2011planck, hincks2010atacama}. In addition, cross-correlations between tSZ, galaxy clustering, and weak lensing are useful for studying the properties of the diffuse gas as well as the mass distribution of galaxy clusters \citep{hojjati2017cross, makiya2018joint, koukoufilippas2020tomographic}. In order to probe intergalactic hot gas, the tSZ effect has several advantages over the X-ray emission originating from Bremsstrahlung. Firstly, its amplitude, characterized by the Compton $y$ parameter, does not depend on the cluster redshift, while X-ray surface brightness scales with $(1+z)^{-4}$, which makes tSZ sensitive to higher redshifts. Secondly, the $y$ parameter depends linearly on the density of gas particles, while X-ray brightness has a quadratic dependence. The X-ray emission is thus more affected by the clumpiness of gas. In addition, the characteristic frequency dependence of tSZ makes it possible to be fully extracted against other sources of radiation such as CMB, Galactic dust thermal emission, and synchrotron emission \citep{remazeilles2011cmb}, while X-ray spectra highly depend on the temperature and composition of sources.

\citet{makiya2018joint}, \citet{pandey2019constraints}, and \citet{koukoufilippas2020tomographic} report measurements of $gy$ with galaxy data from the 2MASS photometric redshift survey and WISE$\times$SuperCOSMOS, the Dark Energy Survey redMaGiC sample, and the 2MASS redshift survey, respectively. In our study, we use the galaxy sample from the fourth data release of the Kilo-Degree Survey (KiDS) \citep{Kuijken_2019} and the Compton $y$ map from the 2015 data release of the {\textit{Planck}} mission \citep{aghanim2016planck}. The previous studies have larger sky coverage but lower survey depth, while KiDS covers only about 2\% of the sky but goes as deep as $z\sim 1$. \citet{chiang2020cosmic} has also measured the tSZ signal up to $z\sim 1$, but uses a quasar catalogue at high redshift. Quasars might have a strong feedback effect in the cross-correlation, which is difficult to model. In contrast, our measurement uses a pure galaxy sample that goes to the highest redshift to date.

The galaxy distribution is a biased tracer of the mass distribution, and when used in cross-correlation studies, the galaxy bias is degenerate with the bias of the other tracer. For example, in the $gy$ measurements, the galaxy bias and the gas pressure bias are degenerate. In \citet{koukoufilippas2020tomographic}, \citet{pandey2019constraints}, and \citet{makiya2018joint}, this problem was addressed by measuring the galaxy bias from the galaxy auto-correlation function, which requires careful modelling of the auto-correlation noise. For KiDS, the field-to-field depth variation is large, which makes galaxy auto-correlations challenging to model with precision. We therefore adopt an alternative approach, which consists of measuring the galaxy bias using the CMB lensing convergence as the mass proxy, via the `$g\kappa$' cross-correlation, which is also adopted in studies such as \citet{Ferraro_2015}. The gravitational lensing effect of CMB photons \citep{lewis2006weak} is an unbiased mass tracer of LSS that has been used to cross-correlate with other tracers to study mass clustering and galaxy bias \citep{bianchini2015cross, singh2017cross, hurier2019first}. In this work, we measure the cross-correlation between galaxy positions and the {\textit{Planck}} CMB lensing map \citep{aghanim2018planck} to independently constrain the galaxy bias and to eliminate the need for modelling the galaxy auto-correlation function. It should be noted that the noise modelling of the galaxy distribution also affects the cross-correlations, but only at the covariance level. In this study, we focus on the linear scale properties of the gas and galaxy position while modelling the cross-correlations on non-linear scales with simple one-parameter models. We do not attempt to extract any cosmological information from them. However, with future improvements in data quality, this approach could in principle be generalised to probe non-linear scales.

We note that the galaxy bias could also be constrained from the cross-correlation between foreground galaxy positions and background galaxy shear, known as galaxy-galaxy lensing. However, at high precision, the interpretation of galaxy lensing requires the modelling of non-lensing effects such as the source-lens clustering and the intrinsic alignments \citep{Hamana_2002, hall2014intrinsic, Valageas_2013} and shape measurement residual systematics. These are extensively studied in their own right, but in this work we intend to highlight the feasibility of using the galaxy distribution as a proxy for the mass distribution with CMB lensing as the galaxy-mass calibration tool. Although CMB lensing has a much lower signal-to-noise than galaxy lensing for a given sky area, it extends to much higher redshift and is immune to most of the non-lensing effects that can contaminate galaxy lensing.

This paper is structured as follows: In Sect.  \ref{sect:model} we describe the theoretical model we use for the cross-correlations; Section \ref{sect:data} introduces the dataset that we are using; Section \ref{sect:measurements} presents the method to measure cross-correlations, as well as our estimation of covariance matrix, likelihood, and systematics; Section \ref{sect:results} presents the results; Section \ref{sect:discussions} discusses the results and summarises our conclusion. Throughout this study, we assume a flat $\Lambda$CDM cosmology with fixed cosmological parameters from \citet{planckcosmo18} as our fiducial cosmology: $ (h,\Omega_\mathrm{c} h^2,\Omega_\mathrm{b} h^2, \sigma_8,n_\mathrm{s}) = (0.676, 0.119, 0.022, 0.81, 0.967)$. The impacts of fixing cosmological parameters are discussed in Section. \ref{sect:discussions}.

\section{Models}
\label{sect:model}

We measure the angular cross-correlation in harmonic space. In general, the angular cross-correlation between two projected tracers, $u$ and $v$, at scales $\ell \gtrsim 10$ are well computed by the Limber approximation \citep{limber1953analysis, kaiser1992weak}:

\begin{equation}
    C_{\ell}^{uv}  =  \int _0 ^{\chi_{\mathrm{H}}} \frac{\mathrm{d}\chi}{\chi^2}W^u(\chi)W^v(\chi)P_{UV}\left(k=\frac{\ell+1/2}{\chi}, z(\chi)\right), 
\end{equation}
where $\chi$ denotes the comoving distance; $\chi_{\mathrm{H}}$ is the comoving distance to the horizon; $W^u(\chi)$ is the radial kernel of tracer $u$; $P_{UV}(k, z)$ is the three-dimensional (3D) cross-power spectrum of associated 3D fluctuating tracers $U$ and $V$:

\begin{equation}
    \left\langle \delta_U(\boldsymbol{k})\delta_V(\boldsymbol{k}') \right\rangle = (2\pi)^3\delta(\boldsymbol{k}+\boldsymbol{k}')P_{UV}(k).
\end{equation}

In this work, the fluctuating physical quantity for tSZ is the 3D electron pressure fluctuation $\Delta P_{\mathrm{e}}$; for galaxy number counts it is the 3D galaxy overdensity $\delta_\mathrm{G}$; for CMB lensing it is the 3D mass overdensity $\delta_{\mathrm{M}}$. We note that, throughout this paper, the two-dimensional (2D) projected tracers (projected galaxy number, Compton $y$ and lensing convergence $\kappa$) are labelled as lower case letters $g$, $y$, and $\kappa$; while the corresponding 3D tracer (galaxy number distribution, electron pressure, and mass distribution) are labelled as capital letters $G$, $P$, and $M$.

In this work we measure $C^{\mathrm{g}y}_{\ell}$ and $C^{\mathrm{g}\kappa}_{\ell}$. The angular fluctuation of galaxy number density $\Delta_g$ is the 2D projection of 3D number-density fluctuations:
\begin{equation}
\Delta_{\mathrm{g}}(\hat{\boldsymbol{\theta}})=
\int_0 ^{\chi_{\mathrm{H}}} d \chi \frac{H(z)}{c}n_{\mathrm{g}}(z) \delta_{\mathrm{G}}(\chi(z) \hat{\boldsymbol{\theta}}, z)
,\end{equation}
where $c$ is the speed of light; $H(z)$ is the Hubble constant at redshift $z$; $\delta_{\mathrm{G}}(\chi(z) \hat{\boldsymbol{\theta}}, z)$ is the 3D galaxy number density fluctuation, and $n_{\mathrm{g}}(z)$ is the normalised redshift distribution of galaxies, which depends on the sky survey. At large scales, we model the number density fluctuations so that it is proportional to the underlying mass fluctuation: $\delta_{\mathrm{G}}= \bg\delta_{\mathrm{M}}$ where $\bg$ is the mean galaxy bias of the galaxy population and $\delta_{\mathrm{M}}$ is the total mass overdensity. The galaxy kernel is given as
\begin{equation}
W^{\mathrm{g}}(\chi) = \frac{H(\chi)}{c} n_{\mathrm{g}}[z(\chi)]
\label{eq:wg}
.\end{equation}

The tSZ signal is parametrised by the Compton-$y$ parameter, given by:
\begin{equation}
    y(\hat{\boldsymbol{\theta}})=\frac{\sigma_{\mathrm{T}}}{m_{e} c^{2}} \int_0 ^{\chi_{\mathrm{H}}} \frac{\rm{d} \chi}{1+z} P_{\mathrm{e}}(\chi \hat{\boldsymbol{\theta}})
,\end{equation}
where $\sigma_{\mathrm{T}}$ is the Thomson scattering cross-section; and $m_{\mathrm{e}}$ is the  electron mass; $P_{\mathrm{e}}$ is the gas electron pressure. Like the galaxy number density fluctuation, we also model the intergalactic gas overdensity as a linearly biased tracer of the underlying mass fluctuation at large scales \citep{goldberg1999microwave, van2014detection}, so the pressure fluctuation $\Delta P_{\mathrm{e}}\equiv P_{\mathrm{e}}-\left\langle P_{\mathrm{e}} \right\rangle= \left\langle P_{\mathrm{e}} \right\rangle \delta_{\mathrm{gas}}= \left\langle b_{y} P_{\mathrm{e}} \right\rangle\delta_\mathrm{m}$, where $\delta_{\mathrm{gas}}$ denotes the gas overdensity; $b_{y}$ is the gas bias; and $\left\langle P_{\mathrm{e}} \right\rangle\equiv \left\langle n_{\mathrm{e}}k_\mathrm{B}T_{\mathrm{e}}\right\rangle$ is the mean electron pressure in gas halos. The combination `gas pressure bias' $\bpe$ is the mean pressure weighted by gas bias, which is related to the thermodynamics of gas inside halos. The tSZ kernel is given by 
\begin{equation}
\label{eq:wy}
    W^{\mathrm{y}}(\chi)=\frac{ \sigma_{\mathrm{T}}}{m_{\mathrm{e}} c^{2}}\frac{ 1}{1+z(\chi)}
.\end{equation}

The CMB lensing kernel is the mass fluctuation convolved with the lensing kernel:
\begin{equation}
    \label{eq:wkappa}
    W^{\mathrm{\kappa}}(\chi)=\frac{3 H_{0}^{2} \Omega_{m}}{2 a c^2} \chi \frac{\chi_{\mathrm{CMB}}-\chi}{\chi_{\mathrm{CMB}}}
,\end{equation}
where $\chi_{\mathrm{CMB}}$ is the comoving distance to the last-scattering surface at $z\sim 1100$; $a$ denotes the scale factor. Therefore, in the linear regime, all three tracers are modelled as linearly biased mass fluctuation convolved with respective kernels, so the linear cross-power spectrum $P_{UV}(k)$ is the linearly biased matter power spectrum:

\begin{equation}
\begin{aligned}
    P^{\mathrm{lin}}_{\mathrm{GP}}(k) &= \bg\bpe P^{\mathrm{lin}}(k)\\
    P^{\mathrm{lin}}_{\mathrm{GM}}(k) &= \bg P^{\mathrm{lin}}(k),
\end{aligned}
\end{equation}
where $P^{\mathrm{lin}}(k)$ is the linear matter power spectrum. For simplicity, hereafter we omit the redshift dependence in the notation of $\bg$ and $\bpe$.

Following the method given in \citet{hang2020galaxy}, to account for non-linear effects at small scales, we model the non-linear portion of our power spectra as an unknown amplitude multiplied by a physical model template:
\begin{equation}
\begin{aligned}
    P^{\mathrm{nl}}_{\mathrm{GP}}(k) &= \cgy T^{\mathrm{nl}}_{\mathrm{GP}}(k),\\
    P^{\mathrm{nl}}_{\mathrm{GM}}(k) &= \cgk T^{\mathrm{nl}}_{\mathrm{GM}}(k),
\end{aligned}
\end{equation}
 where $T^{\mathrm{nl}}(k)$ is the additional non-linear templates. $\cgy$ and $\cgk$ are two re-scaling parameters that account for differences between the amplitudes of non-linear $gy$, $g \kappa$, and dark matter cross-correlations. The total power spectrum is then modelled as: 
\begin{equation}
\begin{aligned}
    P_{\mathrm{GP}}(k)&=\bg\bpe P^{\mathrm{lin}}(k)+\cgy T^\mathrm{nl}_{\mathrm{GP}}(k), \\
    P_{\mathrm{GM}}(k)&=\bg P^{\mathrm{lin}}(k)+\cgk T^\mathrm{nl}_{\mathrm{GM}}(k). \\
\end{aligned}
\label{eq:fullmps}
\end{equation}

\revised{In the standard halo model for non-linear matter power spectrum, the two-halo (linear part) and one-halo (non-linear part) terms both depend on profile amplitudes, so their amplitudes should be correlated. However, there might be some uncertainties in the modelling that might break this correlation, for example, the uncertainty in halo mass function. In addition, \citet{koukoufilippas2020tomographic} points out that the profiles of tracers in Fourier space may not be independent, so the authors introduce a free parameter $\rho_{y\mathrm{g}}$ for the 1-halo term to account for it. In our study, $c_{\mathrm{g}y}$ and $c_{\mathrm{g}\kappa}$ account for combinations of such uncertainties to the first order. }

To ensure that our constraints on linear bias are robust to the exact non-linear models, we try three well-used models for the non-linear power spectrum templates as well as trying a purely linear model: 

1) The first is the halo model: the non-linear power spectrum is given by the one-halo term of the halo model \citep{COORAY_2002, Seljak_2000}:
\begin{equation}
\begin{aligned}
T^{\mathrm{nl}}_{UV}(k) &=P^{\mathrm{1h}}_{UV}(k)\equiv\int_0^{\infty} \dr M \frac{\dr n}{\dr M} p_U(k \mid M) p_V(k \mid M),\\
\end{aligned}
\end{equation}
where $\dr n/\dr M$ is the halo mass function and $p_U(k \mid M)$ is the profile of the tracer $U$ with mass $M$ in Fourier space:
\begin{equation}
p_U(k \mid M) \equiv 4 \pi \int_{0}^{\infty} \dr r r^{2} \frac{\sin (k r)}{k r} p_U(r \mid M)\ .
\end{equation}

For CMB lensing, we take the profile that enters the halo model as the dark matter halo profile, which is typically modelled via the Navarro-Frenk-White profile \citep[NFW; ][]{Navarro_1996}:
\begin{equation}
p_{\mathrm{M}}(r \mid M) = \rho_{\mathrm{NFW}}(r \mid M) \propto \frac{1}{r / r_{\mathrm{s}}\left(1+r / r_{\mathrm{s}}\right)^{2}},
\end{equation}
where $r_{\mathrm{s}}$ is the characteristic radius of a dark matter halo, which relates to the halo mass that we take from the mass-concentration relation.

The galaxy population in a halo is divided into centrals and satellites, the abundances of which we relate to halo mass via a Halo Occupation Distribution (HOD) model \citep{Zheng_2005,Peacock_2000}:

\begin{equation}
\begin{aligned}
     N_{\mathrm{c}}(M) &= \frac{1}{2}\left[1+\operatorname{erf}\left(\frac{\log \left(M / M_{\min }\right)}{\sigma_{\ln M}}\right)\right] \\
     N_{\mathrm{s}}(M) &= N_{\mathrm{c}}(M) \Theta\left(M-M_{0}\right)\left(\frac{M-M_{0}}{M_{1}}\right)^{\alpha_{\mathrm{s}}},
\end{aligned}
\end{equation}
where $N_{\mathrm{c}}(M)$ and $ N_{\mathrm{s}}(M)$ are the mean number of central and satellite galaxies respectively. $M_1, M_0, M_{\mathrm{min}}, \sigma_{\mathrm{M}}, \alpha_s$ are free parameters in principle. The galaxy density profile is then:
\begin{equation}
    p_{\mathrm{g}}(k\mid M)=\bar{n}_{\mathrm{g}}^{-1}\left[ N_{\mathrm{c}}(M)+ N_{\mathrm{s}}(M) p_{\mathrm{s}}(k \mid M)\right],
\end{equation}
where $\bar{n}_{\mathrm{g}}$ is the mean galaxy number density. We assume that central galaxies exist at the halo centre while satellites follow the underlying matter distribution, so the satellite profile $p_{\mathrm{s}}$ is the NFW profile. In this work, we fix HOD parameters to $\{\sigma_{\mathrm{M}}, \alpha_s, \log_{10}M_1, \log_{10}M_0, \log_{10}M_{\mathrm{min}}\}=\{0.15, 1, 13, 11.86, 11.68\}$ as constrained from \citet{Zheng_2005}. Here masses are in the unit of $h^{-1}M_{\odot}$. While this may not be a correct description of our galaxy population, we see later that our conclusions are unaffected by the details of our non-linear model.

The $y$ signal derives via the electron pressure profile $p_{\mathrm{e}}(r, M, z)$, which we take from \citet{Arnaud_2010}:

\begin{equation}\begin{aligned}
p_{\mathrm{e}}(r, M, z) &=1.65(h / 0.7)^{2}\, \mathrm{eV} \mathrm{cm}^{-3} \\
& \times E^{8 / 3}(z)\left[\frac{M}{3 \times 10^{14}(0.7 / h) \mathrm{M}_{\odot}}\right]^{2 / 3+\alpha_{\mathrm{p}}} p(x),
\end{aligned}
\label{eq:gnfw}
\end{equation}
where $x \equiv r / r_{500}$ and $E(z) \equiv H(z) / H_{0}$. $r_{500}$ is the radius that encloses a region with average density equal to 500 times the critical density of the Universe. The parameter $\alpha_{\mathrm{p}}=0.12$ as given by \citet{Arnaud_2010}. The self-similar part of the pressure profile $p(x)$ is given by \citep{Nagai_2007}:

\begin{equation}p(x) \equiv \frac{P_{0}(0.7 / h)^{3 / 2}}{\left(c_{500} x\right)^{\gamma}\left[1+\left(c_{500} x\right)^{\alpha}\right]^{(\beta-\gamma) / \alpha}}.\end{equation}
The parameters in $p(x)$ are taken as the best-fitted values from \citet{2013pl}: $\{P_{0}, c_{500}, \alpha, \beta, \gamma\}=\{6.41,1.81,1.33,4.13,0.31\}$.

We note that, in halo model, the gas pressure bias can be expressed as:

\begin{equation}
\left\langle b_{y} P_{\mathrm{e}}\right\rangle(z)=\int_0^{\infty} \dr M \frac{\dr n}{\dr M}(z) b_{\mathrm{h}}(M,z) \int_0^{\infty} \dr r 4 \pi r^{2} p_{\mathrm{e}}(r, M, z)
\label{eq:hmbpe}
,\end{equation}
where 
\begin{equation}
\int_0^{\infty} \dr r 4 \pi r^{2} p_{\mathrm{e}}(r, M, z) \equiv E_{\mathrm{T}}(M, z)
,\end{equation}
which means that
\begin{equation}
\left\langle b_{y} P_{\mathrm{e}}\right\rangle(z)=\int_0^{\infty} \dr M \frac{\dr n}{\dr M}(z) b_{\mathrm{h}}(M,z) E_{\mathrm{T}}(M, z)
.
\label{eq:bpe_gnfw}
\end{equation}
Therefore, the gas pressure bias directly links to the thermal energy of dark matter halos. 

We took the halo mass function and halo bias needed in the halo model from the fitting formulae of \citet{Tinker_2008} and \citet{Tinker_2010}, respectively. We chose this halo model as our fiducial model.

2) \halofit non-linear model: we isolate the purely non-linear part of the \halofit model \citep{Smith_2003, takahashi2012revising} by taking the full model and subtracting linear theory:

\begin{equation}
    T^\mathrm{nl}(k) = P^{\text {HF}}(k) - P^{\text {lin}}(k),
\end{equation}
where $P^{\text {HF}}(k)$ is the \halofit matter power spectrum. While the \halofit model was calibrated to the matter power spectrum only, we hope that the non-linear shape is general enough to capture the correct shape of the galaxy--Pressure and galaxy--matter power spectra that are relevant to our cross-correlations. Any amplitude differences will be absorbed by our multiplicative non-linear coefficients. This non-linear template is also used in \citet{hang2020galaxy}. 

3) Constant non-linear model: The non-linear power spectra are constants:

\begin{equation}
    T^{\mathrm{nl}}_{UV}(k)=P^{\mathrm{1h}}_{UV}(0),
\end{equation}
where $P^{\mathrm{1h}}_{UV}(0)$ is the one-halo term of the halo-model power spectrum at $k=0\,\mathrm{Mpc}^{-1}$.This model should only work on large scales  where the one-halo region can be treated as a point source.

To make sure that our non-linear models are not sensitive to the precise shape of the non-linear power spectrum, we only fit $C_{\ell}$'s in each tomographic bin within an angular scale corresponding to $k<0.7\, \mathrm{Mpc}^{-1}$ via the Limber approximation. In addition, \citet{mead2020including} points out that the halo model is not accurate in the transition between the one- and two-halo regions. To attempt to correct for this, we follow \citet{koukoufilippas2020tomographic} and multiply the power spectra given by 1) and 3) with a scale-dependent quantity:

\begin{equation}
R(k) \equiv \frac{P^{\text {HF}}(k)}{P^{\text {hm}}_{\mathrm{MM}}(k)},
\label{eq:rkdef}
\end{equation}
where $P^\mathrm{HF}(k)$ is from \halofit and $P^{\text {hm}}_{\mathrm{MM}}(k)$ is the matter--matter power evaluated via the halo model (with NFW profiles). Although \cite{mead2020including} showed that the correction required to the halo model in the transition region is not universal, and instead depends on the tracers being modelled, it was shown that attempting some correction is better than no correction at all.

Previous studies on galaxy or gas cross-correlations such as \citet{van2014detection, bianchini2015cross, Kuntz_2015} treat the galaxy or gas distribution to be proportional to mass distribution on all scales, which means they only have two free parameters: $\bg$ or $\bpe$. However, it should be noted that this model is not physically accurate because the galaxy and gas distributions have significantly different non-linear details compared to the matter distribution. \citet{pandey2019constraints} use a linear bias model for the galaxy distribution over the scales $R\gtrsim10$ Mpc, to which our measurements are not sensitive. However, as is indicated in \citet{Sugiyama_2020}, to apply a fully linear model, one needs a scale cut of at least 12 Mpc. The free prefactors for our three models account for different non-linear amplitudes, but our models are probably still not accurate for the details of the non-linear shape. To model the shapes more accurately, one would need to constrain HOD and pressure profile parameters as well as to significantly improve the treatment of the transition region, which is not feasible in this analysis given the noise level of current data. We leave this to future study. In summary, with fixed cosmological parameters, by measuring $C_{\ell}^{\mathrm{g}y}$ and $C_{\ell}^{\mathrm{g}\kappa}$ we can independently constrain and compare $\bg$, $\bpe$, $\cgy$ and $\cgk$ with the three non-linear models introduced above. 

\section{Data}
\label{sect:data}
\subsection{KiDS Data}
We use the lensing catalogue provided by the fourth data release of the KiDS \citep{Kuijken_2019} as our galaxy sample. KiDS is a sky-survey project, which measures the positions and shapes of galaxies using the VLT Survey Telescope (VST) at the European Southern Observatory (ESO). It is primarily designed for weak-lensing applications. The footprint of KiDS DR4 (also called KiDS-1000) is divided into a northern and southern patch, with total coverage of 1006 $\mathrm{deg}^2$ of the sky (corresponding to a fraction of $f_{\mathrm{sky}}=2.2\%$.) The footprint is shown in the upper panel of Fig. \ref{fig:masks}. High-quality images are produced with VST-OmegaCAM. Combining with the VISTA Kilo-degree INfrared Galaxy survey (VIKING; \citealt{2013Msngr.154...32E}), the observed galaxies are photometrically measured in nine optical and near-infrared bands $ugriZYJHK_{\mathrm{s}}$. The KiDS survey covers redshifts $z\lesssim 1.5$, which makes it a useful dataset to trace the history of different components of the LSS into the early Universe. For each galaxy in the lensing catalogue, the ellipticities are measured with the \textsc{lensfit} algorithm \citep{miller2013bayesian}. We only use the `gold subsample' \citep{Wright_2020} of the lensing catalogue since the redshift distribution is more accurately calibrated in this subsample. We present the information of the galaxy sample that we use in Table. \ref{table:tomoinfo}. Note, however, that we do not use the shape information in our fiducial analysis. In Appendix \ref{sec:ggl} we use the shape information to replace CMB lensing as an alternative measurement and sanity check.

\begin{figure}
    \centering
    \includegraphics[width=\columnwidth]{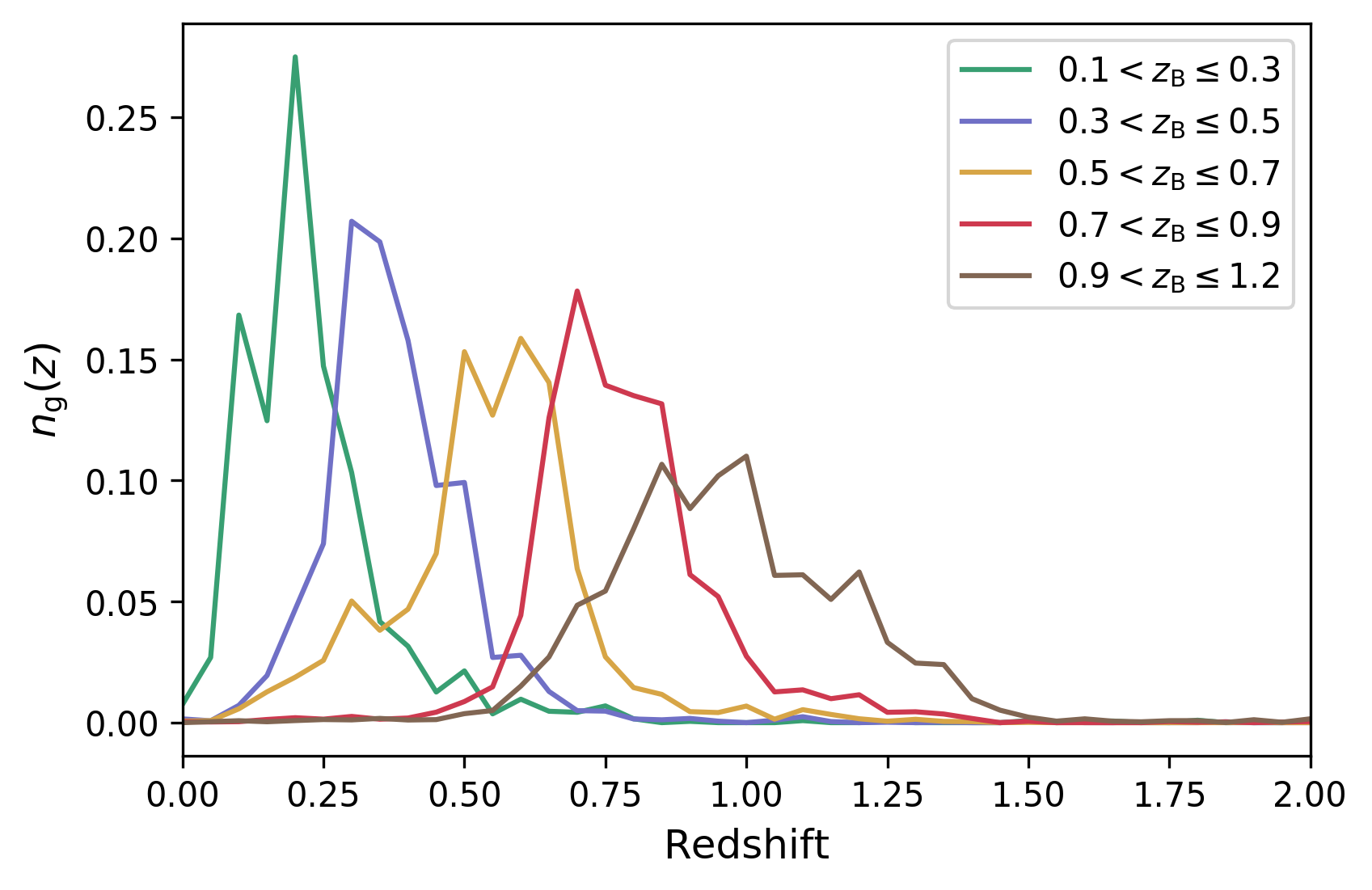}
    \caption{Redshift distributions of the five tomographic bins of the KiDS gold galaxy sample.}
    \label{fig:dndz}
\end{figure}

\begin{table}
\begin{tabular}{llll}\toprule
Bin & $z_{\mathrm{B}}$ range  & Mean redshift & $\bar{n}$ \\\midrule
1   & {(}0.1, 0.3] & 0.23          & 5.73    \\
2   & {(}0.3, 0.5] & 0.38          & 11.87   \\
3   & {(}0.5, 0.7] & 0.54          & 20.18   \\
4   & {(}0.7, 0.9] & 0.77          & 14.81   \\
5   & {(}0.9, 1.2] & 0.96          & 17.20   \\\bottomrule
\end{tabular}
\caption{Information on the KiDS galaxy sample in each tomographic bin. $\bar{n}$ stands for mean galaxy number in a \healpix pixel with $\textsc{nside}=1024$.}
\label{table:tomoinfo}
\end{table}
We perform a tomographic measurement of cross-correlations by dividing the galaxy catalogue into five redshift bins according to the best-fit photometric redshift $z_{\mathrm{B}}$ of each galaxy. These are the same redshift bins used in the KiDS-1000 cosmology papers \citep{asgari2020kids1000, heymans2020kids1000, troester2020kids1000}. The redshift distribution of each bin is calibrated using Self-Organising Maps (SOM) as described in \citet{2020A&A...637A.100W, hildebr2020kids1000}. We note that the SOM-calibrated redshift distributions in this study are not exactly the same as \citet{hildebr2020kids1000} in which the redshift distributions are calibrated with a galaxy sample weighted by the \textit{lensfit} weight, while in this work the redshift distributions are calibrated with the raw, unweighted sample. The redshift distributions of the 5 tomographic bins are shown in Fig. \ref{fig:dndz}. Galaxy overdensity maps are produced for each tomographic bin in the \healpix \citep{Gorski_2005} format with $\textsc{nside} = 1024$, corresponding to a pixel size of 3.4 arcmin. For each tomographic bin, the galaxy overdensity in each pixel is given as
\begin{equation}
    \Delta_{\mathrm{g}, i} = \frac{n_i-\bar{n}}{\bar{n}}
,\end{equation}
where $i$ denotes the pixel index, $n_i$ is the number of galaxies in the $i$-th pixel and $\bar{n}$ is the average galaxy number of all the pixels in the footprint and the given redshift bin. The galaxy mask for the cross-correlation measurement is just the KiDS footprint, which is presented in the upper panel of Fig. \ref{fig:masks}.

\subsection{tSZ data}

\begin{figure}
    \centering
    \includegraphics[width=\columnwidth]{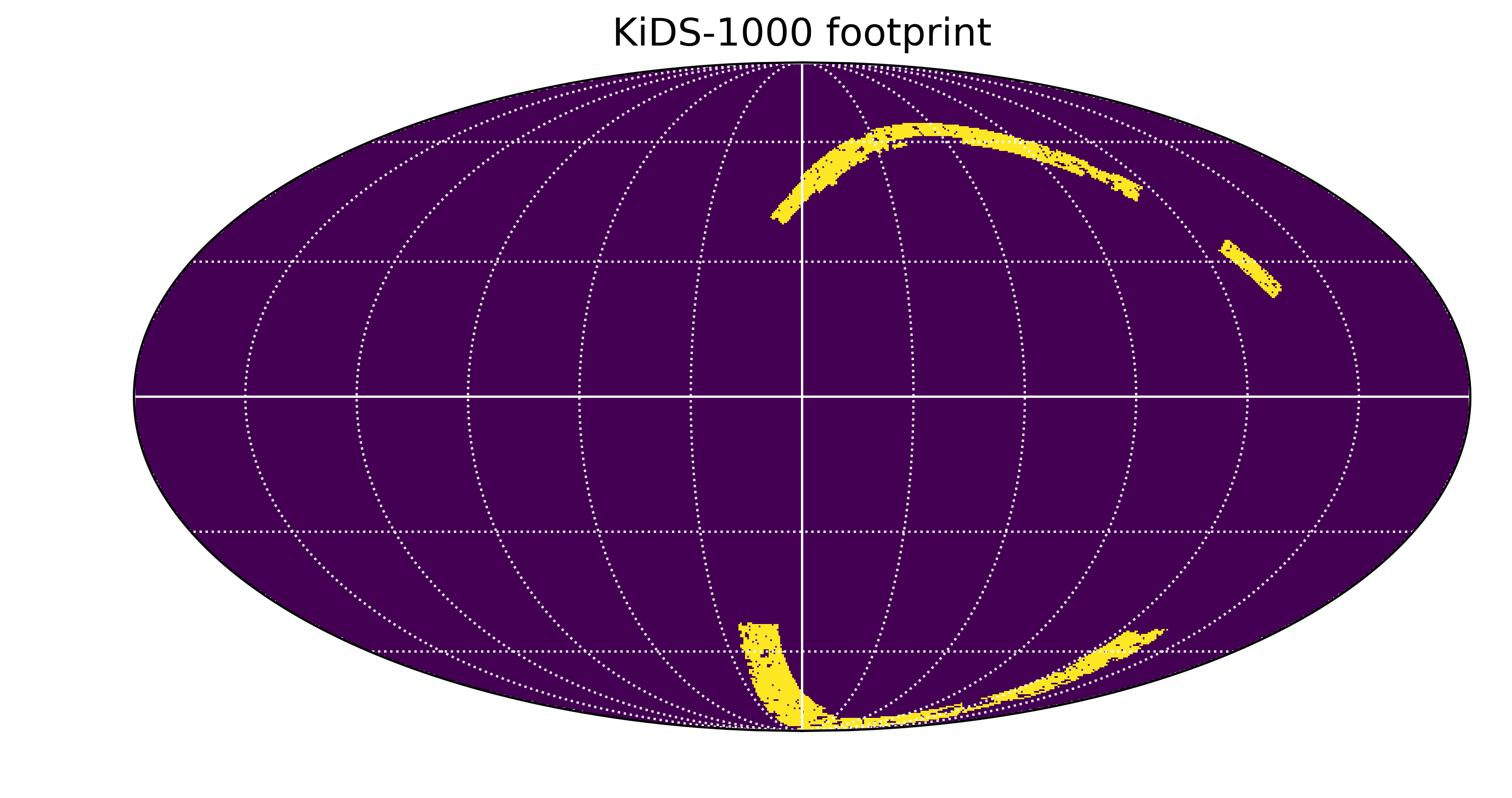}
    \includegraphics[width=\columnwidth]{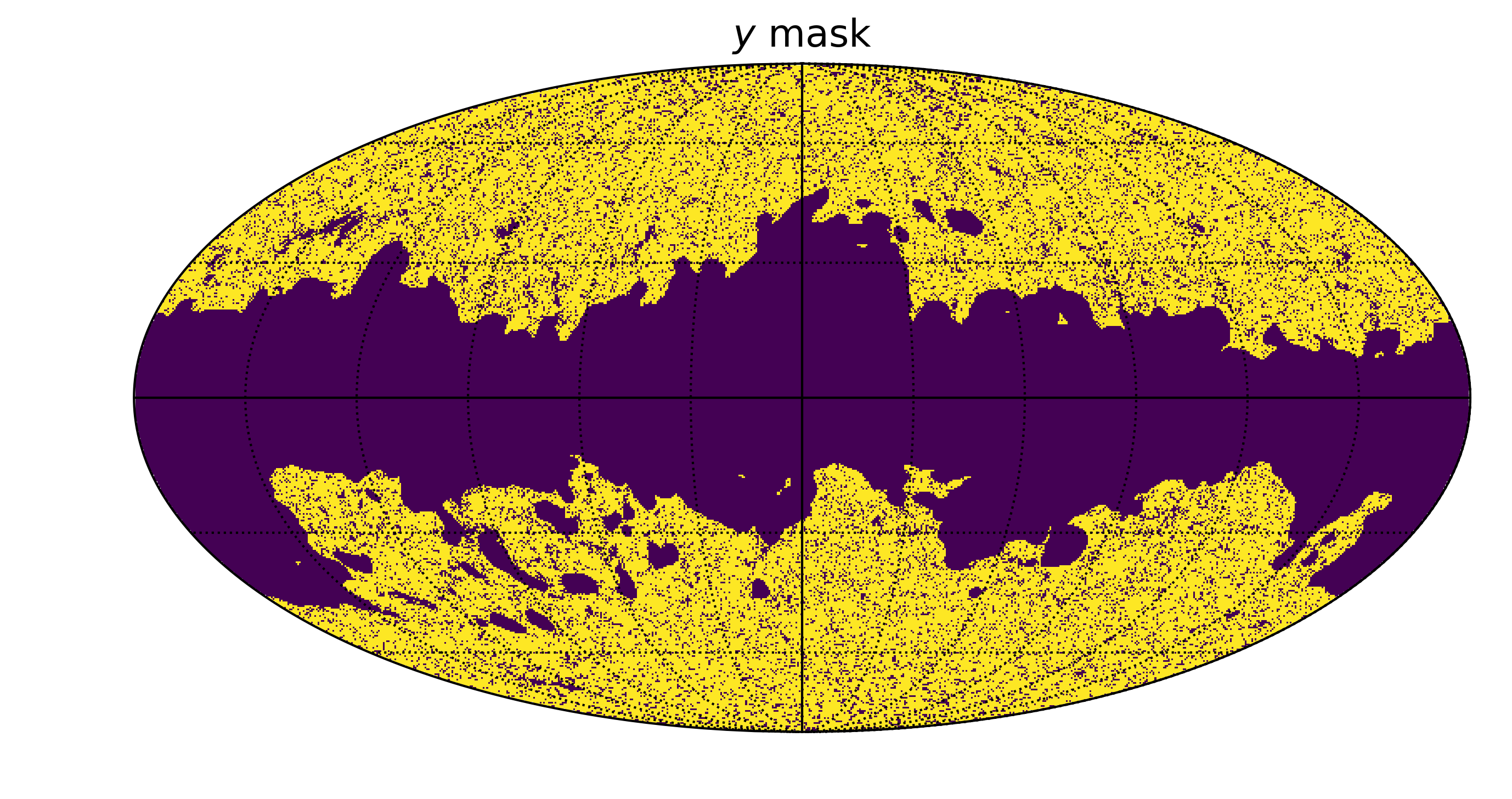}
    \includegraphics[width=\columnwidth]{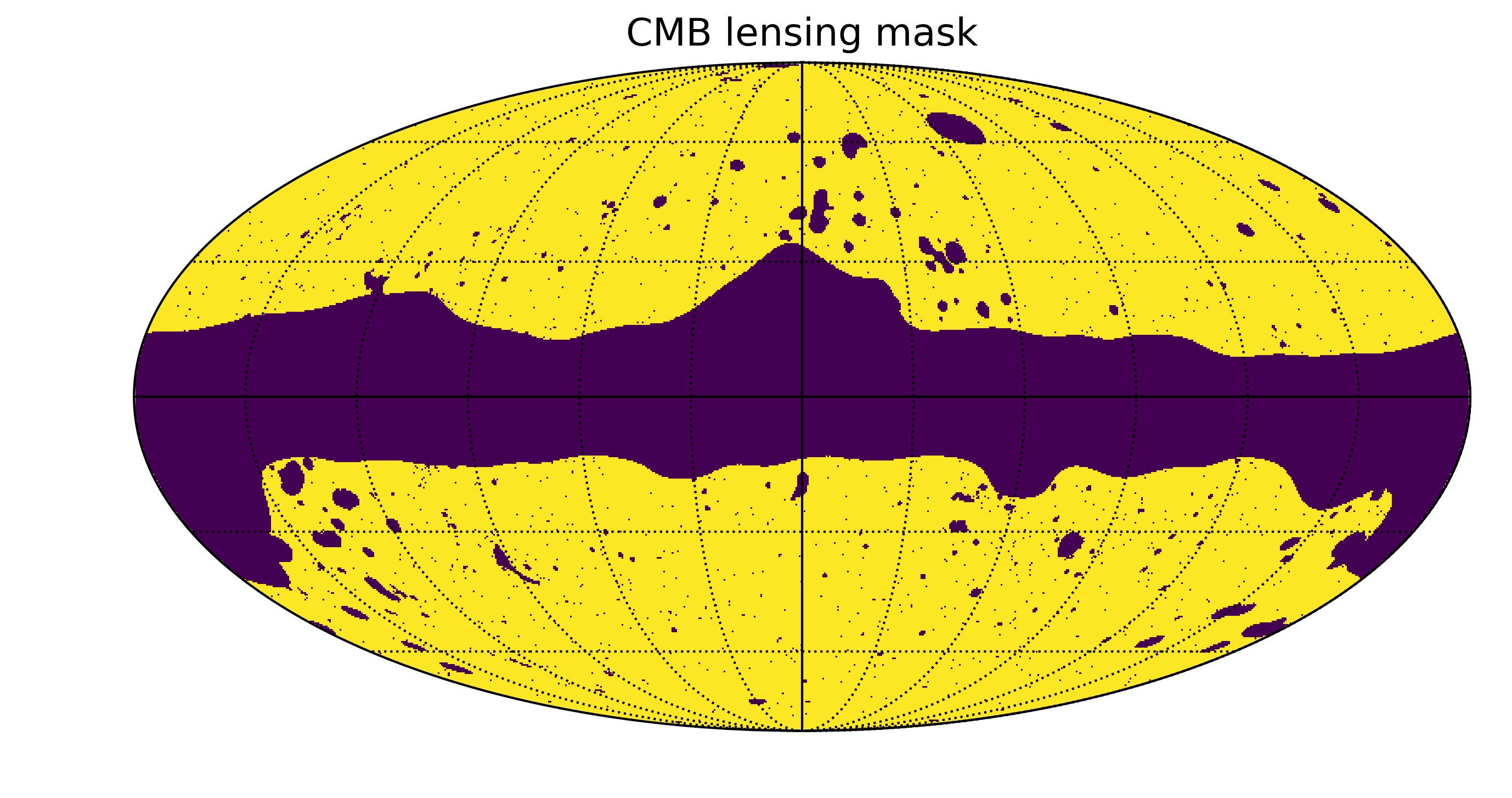}
    \caption{KiDS-1000 footprint and masks for the {\textit{Planck}} $y$ map and CMB lensing map. Regions in purple are masked out.}
    \label{fig:masks}
\end{figure}

We use the all-sky Compton-$y$ map presented in \citet{aghanim2016planck} for the tSZ signal. The $y$ map that we used is constructed with the Modified Internal Linear Combination Algorithm (MILCA; \citealt{Hurier_2013}), which properly suppresses scale-dependent contamination, and projects out the CMB signal. {\textit{Planck}} has also released another $y$ map constructed with the Needlet Internal Linear Combination (NILC; \citealt{Basak:2011yt}) method. Although the MILCA and NILC $y$ maps agree with each other in most of the relevant studies, the NILC map turns out to be noisier \citep{aghanim2016planck}. Therefore, we apply the {\textit{Planck}} MILCA map in this study\revised{ and take the NILC $y$ map as a consistency check}. Both $y$ maps have a beam full width at half Maximum (FWHM) of 10 arcmin.

Before calculating the $gy$ cross-correlation, we mask out the Milky Way and point sources with a joint mask of the {\textit{Planck}} 60\% Galactic mask and point source mask. The combined mask is shown in the middle panel of Fig. \ref{fig:masks}. The mask and $y$ map are originally provided in the \healpix format with $\textsc{nside}=2048$ and we degrade them into $\textsc{nside}=1024$ to match the resolution of the KiDS galaxy overdensity map.

To evaluate the CIB contamination in the galaxy-tSZ cross-correlation, we also introduced the {\textit{Planck}} 545 GHz CIB intensity map as a CIB template \citep{2016planckcib}. The CIB intensity map was generated with the generalised-NILC \citep[GNILC]{Remazeilles_2011} method and has an angular resolution of 5 arcmin. We first convolved the CIB map with a $\sqrt{10^2-5^2}=8.66$ arcmin Gaussian filter to match its resolution to the 10 arcmin of the y map before degrading the map to $\textsc{nside}=1024$.

\subsection{CMB lensing data}

We used the \revised{tSZ-deprojected} {\textit{Planck}} CMB lensing map from the 2018 Data Release \citep{aghanim2018planck} to measure the galaxy-CMB Lensing cross-correlation. The map is provided in the format of the spherical harmonic transformation of the lensing convergence $\kappa_{\ell m}$, which is related to the lensing potential $\phi$ via

\begin{equation}
    \kappa_{\ell m} = \frac{\ell(\ell+1)}{2}\phi_{\ell m}
.\end{equation}

\revisednew{There might be reconstruction bias in the CMB lensing map, which is typically descriped by a lensing amplitude parameter, $A_{\mathrm{L}}$. \citet{aghanim2018planck} reports a value of $A_{\mathrm{L}}$ that is slightly greater than 1, with a significance of $\sim 2\sigma$. \citet{efstathiou2020detailed} points out that this might be due to the fluctuations in the temperature power spectrum at high $\ell$. In $\mathrm{g}\kappa$ cross-correlation, $A_{\mathrm{L}}$ degenerates with galaxy bias. This analysis cannot break the degeneracy, so we take the prior information, $A_{\mathrm{L}}=1$. }

We first transformed $\kappa_{\ell m}$ back into a \healpix $\kappa$ map with $\textsc{nside}=1024$. The corresponding mask is provided along with the CMB lensing data. It is shown in the lower panel of Fig. \ref{fig:masks}.

\section{Measurements}
\label{sect:measurements}
\subsection{Cross-correlation measurements}
\label{subsect:ccmeasurements}
The cross-correlation between two sky maps, that are smoothed with the beam window function $b_{\mathrm{beam}}(\ell)$, is related to the real $C_{\ell}$ with

\begin{equation}
    \hat{C}^{uv}_{\ell} = C^{uv}_{\ell}b^u_{\mathrm{beam}}(\ell)b^v_{\mathrm{beam}}(\ell)b^u_{\mathrm{pix}}(\ell)b^v_{\mathrm{pix}}(\ell)
    \label{eq:pseudo_cl}
,\end{equation}
where $\hat{C}^{uv}_{\ell}$ denotes the smoothed $C_{\ell}$ between sky map $u$ and $v$; and $b_{\mathrm{pix}}(\ell)$ is the pixelisation window function. In our analysis we take the Gaussian window function which is given by
\begin{equation}
    b_{\mathrm{beam}}(\ell) = \exp \left(-\ell(\ell+1)\sigma^2/2\right) 
,\end{equation}
where $\sigma = \mathrm{FWHM}/{\sqrt{8\ln2}}$. The pixelisation window function corresponding to $\textsc{nside}=1024$ is provided by the \healpix package. We note that for the KiDS galaxy map, FWHM = 0.

We use \textsc{polspice} to estimate the angular cross-power spectra. Mode-coupling due to mask and beam smoothing are corrected during this process. Fourier ringings are reduced by setting the internal parameters of \textsc{polspice} \texttt{apodizesigma}=60 and \texttt{thetamax}=60 deg. The measured angular power spectra are binned into 10 linear bins from $\ell=100$ to $\ell=1100$. The high limit of $\ell$ corresponds to the Planck beam, which has a size of 10 arcmin.

\begin{figure}
    \centering
    \includegraphics[width=0.9\columnwidth]{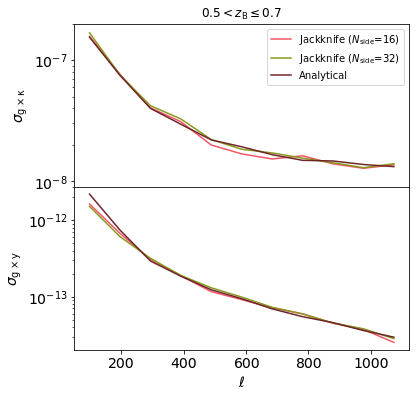}
    \caption{Standard deviation calculated from the diagonal of $\mathrm{Cov}^{\mathrm{JK}}(C^{\mathrm{g}y}_{\ell}, C^{\mathrm{g}y}_{\ell})$ and $\mathrm{Cov}^{\mathrm{ANA}}(C^{\mathrm{g}y}_{\ell}, C^{\mathrm{g}y}_{\ell})$ in the third tomographic bin. We also plot the standard deviation from the jackknife covariance matrix with $N_{\mathrm{side}}=16$ as a consistency check.}
    \label{fig:std_compare}
\end{figure}

\subsection{Covariance matrix}
\begin{figure*}
    \centering
    \includegraphics[width=0.8\textwidth]{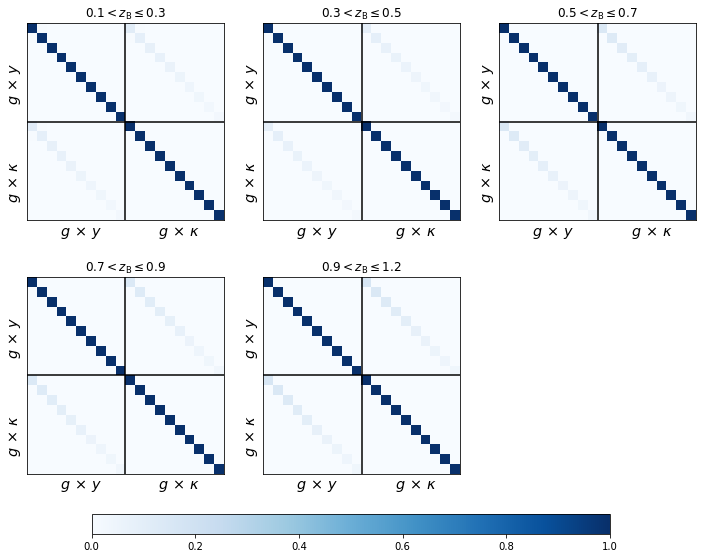}
    \caption{Full correlation coefficient matrices of covariance matrices (defined in Eq.\eqref{eq:covfull}) of the $gy$ and the $g\kappa$ cross-correlations in each tomographic bin. Each covariance matrix consists of four sub-matrices, corresponding to the covariances of $C^{\mathrm{g}y}_{\ell}$ and $C^{\mathrm{g}\kappa}_{\ell}$ (block diagonals), as well as their cross-covariance. For each sub-matrix, the pixels show the cross coefficient between binned $C_{\ell}$'s, where $\ell$ bins are defined in subsection \ref{subsect:ccmeasurements}.}
    \label{fig:covmat}
\end{figure*}
We combined two methods to estimate the covariance matrix in our $C_{\ell}$ measurement: One is jackknife resampling, the other is an analytical Gaussian covariance matrix. For jackknife resampling, we generated 415 jackknife samples by masking out pixels corresponding to $\textsc{nside}=32$ (which has a size of 1.83 degree) in turn from the KiDS galaxy overdensity map. Smaller jackknife pixels would fail to estimate the variance at large scales, while with larger jackknife pixels we would not have enough realisations. So we chose this intermediate jackknife pixel size to balance the pixel size with the sample size. However, as shown in Fig. \ref{fig:std_compare}, the jackknife pixel with $\textsc{nside}=16$ (corresponding to an angular size of 3.66 degree) gives a consistent standard deviation. The cross-correlations $C^{\mathrm{g}y}_{\ell}$ and $C^{\mathrm{g}\kappa}_{\ell}$ are measured with each of these jackknife samples and the covariance matrix is calculated with

\begin{equation}
    \operatorname{Cov}^{\mathrm{JK}}\left(C_{\ell}^{u v}, C_{\ell^{\prime}}^{w z}\right)=\frac{N_{\mathrm{JK}}-1}{N_{\mathrm{JK}}} \sum_{n=1}^{N_{\mathrm{JK}}} \Delta C_{\ell}^{u v,(n)} \Delta C_{\ell^{\prime}}^{w z,(n)}
,\end{equation}
where $N_{\mathrm{JK}}=415$ denotes the number of jackknife samples; $uv, wz\in\{gy, g\kappa\}$; $\Delta C_{\ell}^{u v,(n)}$ is the difference between the cross-correlation of the $n$-th jackknife sample and the mean cross-correlation over all samples.

Since the KiDS footprint is only $\sim 2\%$ of the sky, it is hard to generate enough jackknife samples to fully recover the true covariance matrix; as such, the off-diagonal components of the jackknife covariance matrix are noisy. In addition, since different jackknife regions have slightly different shapes, we could not recover the mode-coupling in the covariance matrix associated with the whole map geometry. To better estimate the off-diagonal components and account for mode-coupling accurately, we also estimated the covariance matrix using an analytical method.

The main contribution to the covariance matrix is from a Gaussian random field:

\begin{equation}\operatorname{Cov}^{\mathrm{G}}\left(C_{\ell}^{u v}, C_{\ell^{\prime}}^{w z}\right)=\delta_{\ell \ell^{\prime}} \frac{C_{\ell}^{u w} C_{\ell^{\prime}}^{v z}+C_{\ell}^{u z} C_{\ell^{\prime}}^{v w}}{f_{\mathrm{sky}}(2 \ell+1)}
\label{eq:covg}
,\end{equation}
where $f_{\mathrm{sky}}$\revised{=2.2\%} is the sky fraction. Sky masks introduce non-zero coupling between different $\ell$. To account for this, we used the method given by \citet{Efstathiou_2004} and \citet{Garc_a_Garc_a_2019} and implemented in the \textsc{namaster} package \citep{Alonso_2019} \footnote{We note that \textsc{namaster} can also measure $C_{\ell}$ and their results agree with \textsc{polspice}, but \textsc{namaster} is significantly slower than \textsc{polspice} when calculating more than 1000 jackknife cross-correlations. So we only use it to calculate the analytical covariance matrix, which \textsc{polspice} cannot do.}. The auto-power spectra in \eqref{eq:covg} are directly measured from maps so that noise auto-spectra can also be included; the cross-power spectra are instead calculated from the theoretical model described in Section~\ref{sect:model}, since their measurements are significantly noisier. 

The non-Gaussian term includes a connected contribution resulting from the small-scale non-linear clustering of the tracers, related to the trispectrum of the tracers. According to \citet{koukoufilippas2020tomographic}, \citet{Barreira_2018}, and \citet{Nicola_2020}, this contribution is only significant for low redshifts $z\lesssim0.2$; therefore, we neglect it in our covariance matrix. Another non-Gaussian contribution is the super-sample covariance \citep[SSC]{Takada_2013} resulting from mode mixing between observed in-survey and the unobserved out-of-survey modes, we also ignore this in this work.

To calculate the analytical covariance matrices, we need to use model parameters that we do not know \textit{a priori}. So we follow \citet{koukoufilippas2020tomographic} and take a two-step fitting: we first take an assumed value for the model parameters to calculate these covariance matrices and then use these to find the best-fit parameters. We then update the covariance matrix, using these best-fit parameters, and fit for the parameters again. The best-fit parameters from this second round of fitting are taken to be our fiducial results.

The diagonal components of the jackknife and analytical covariance matrices generally agree with each other (see Fig. \ref{fig:std_compare} as an example), this justifies that we can ignore the non-Gaussian contribution in the covariance matrix. To ensure that we recover realistic error bar sizes, we combine the variance estimated from the jackknife covariance matrix with the analytical covariance matrix, as in \cite{koukoufilippas2020tomographic}, to account for the coupling between different modes caused by masks and non-Gaussianities while avoiding the statistical noise in the jackknife estimator. Therefore our final covariance matrix is:

\begin{equation}\operatorname{Cov}_{i j}=\operatorname{Cov}_{i j}^{\mathrm{ANA}} \sqrt{\frac{\operatorname{Cov}_{i i}^{\mathrm{JK}} \operatorname{Cov}_{j j}^{\mathrm{JK}}}{\operatorname{Cov}_{i i}^{\text {ANA }} \operatorname{Cov}_{j j}^{\text {ANA }}}}.
\label{eq:covfull}
\end{equation}
The correlation coefficient matrices of all the tomographic bins are shown in Fig.~\ref{fig:covmat}.

\subsection{Likelihood}

Since we are working with a wide $\ell$ range, there are many degrees of freedom in each $\ell$ bin. According to the central limit theorem, the bin-averaged $C_{\ell}$'s obey a Gaussian distribution around their true values. Thus we assume that the measured power spectra follow a Gaussian likelihood:

\begin{equation}
    -2 \ln L(\boldsymbol{D} \mid\vec{q})=\chi^{2} \equiv(\boldsymbol{D}-\boldsymbol{M}(\vec{q}))^{T} \mathrm{Cov}^{-1}(\boldsymbol{D}-\boldsymbol{M}(\vec{q}))
,
\label{eq:like}
\end{equation}
where $\vec{q}\equiv\left\{\bg\times\left\langle b_{y}P_{\mathrm{e}} \right\rangle ,\bg, \cgy, \cgk\right\}$ stands for our model parameters (for the `linear model' we only have two parameters $\vec{q}\equiv\left\{\bg\times\left\langle b_{y}P_{\mathrm{e}} \right\rangle ,\bg \right\}$); the data vector $\boldsymbol{D}\equiv (C^{\mathrm{g}y}_{\ell}, C^{\mathrm{g}\kappa}_{\ell})$ is a concatenation of measured $gy$ and $g\kappa$ cross-correlations; $\boldsymbol{M}(\vec{q})$ is the cross-correlation predicted by model described in Section~\ref{sect:model} with parameter $\vec{q}$. \revised{The likelihoods are calculated separately in each tomographic bin}. We note that the gas pressure bias $\bpe$, which we are primarily interested in, is the ratio between the first two model parameters.

We sample the posterior distribution of model parameters using the Markov chain Monte Carlo method (MCMC) using the \texttt{emcee} package \citep{Foreman_Mackey_2013}. We take flat priors for all four model parameters:
\begin{equation}
\begin{aligned}
    0 &\leq \bg\leq 3, \\
    0 &\leq \bg\times\bpe\leq 9,\\
    0 &\leq \cgk\leq 10, \\ 
    0 &\leq \cgy\leq 10. \\
\end{aligned}
\end{equation}
The lower boundaries of the prior is a physical constrain, indicating that the parameters cannot be negative; the upper boundaries are set so that at least $5\sigma$ of the marginalised posterior distribution falls in these ranges.

\revisednew{We are not doing a joint analysis of all the tomographic bins, so the correlations between different redshift bins are not relevant to our analysis. Following \citet{Crocce_2015}, \citet{koukoufilippas2020tomographic}, and \citet{hang2020galaxy}, our model parameters are fit independently for each of the tomographic bins.} The theoretical model is calculated using the Core Cosmology Library package \citep{Chisari_2019}.

\subsection{Systematics}
\subsubsection{CIB contamination}
The CIB radiation is the accumulated emission from early galaxy populations spanning a large range of redshifts, mostly generated from dust thermal radiation around extragalactic star-formation regions \citep{Hauser_2001}. The tSZ map is contaminated by residual CIB \citep{Hurier_2015, Yan_2019}, which dominates extragalactic signals at high frequency and high redshifts. This residual might contaminate our galaxy-tSZ cross-correlation. We follow the method in \citet{koukoufilippas2020tomographic} to model the CIB contamination in the $y$ map as a factor $\alpha_{\mathrm{CIB}}$ times a CIB template map, which is taken to be the {\textit{Planck}} CIB intensity map at 545 GHz \citep{2016planckcib}:

\begin{equation}
    \hat{y}(\boldsymbol{\theta}) = y(\boldsymbol{\theta}) + \alpha_{\mathrm{CIB}} I_{\mathrm{CIB}}(\boldsymbol{\theta})
,\end{equation}
where $\hat{y}$ denotes the contaminated $y$ map and $I_{\mathrm{CIB}}$ denotes the CIB intensity in 545 GHz. So the measured galaxy-tSZ cross-correlation is given by:
\begin{equation}
    C^{\mathrm{g\hat{y}}}_{\ell} = C^{\mathrm{g}y}_{\ell} + \alpha_{\mathrm{CIB}} C^{\mathrm{g}I_{\mathrm{CIB}}}_{\ell}
,\end{equation}
where the galaxy-CIB cross-correlation $C^{\mathrm{g}I_{\mathrm{CIB}}}_{\ell}$ can be directly measured with the {\textit{Planck}} CIB map. For $\alpha_{\mathrm{CIB}}$, we take the value $\alpha_{\mathrm{CIB}}=(2.3\pm 6.6) \times 10^{-7} (\mathrm{MJy} / \mathrm{sr})^{-1}$ reported in \citet{Alonso_2018}. 

\subsubsection{Cosmic magnification}

The measured galaxy overdensity depends not only on the real galaxy distribution but also on lensing magnification induced by the line-of-sight mass distribution \citep{1989A&A...221..221S, 1989ApJ...339L..53N}. This magnification, or so-called cosmic magnification, has two effects on the measured galaxy overdensity: i) overdensities along the line-of-sight cause the local angular separation between source galaxies to increase, so the galaxy spatial distributions are diluted and cross-correlation is suppressed; ii) lenses along line-of-sight magnify the flux of source galaxies such that fainter galaxies enter the observed sample, so the overdensity increases. These effects bias galaxy-related cross-correlations, especially for high-redshift galaxies \citep{hui2007anisotropic, ziour2008magnification, hilbert2009ray}. To take these two effects into account, we modify the expression for the galaxy overdensity \citep{hui2007anisotropic}:

\begin{equation}
    \hat{\delta}_\mathrm{g}(z) = \bg\delta_\mathrm{m}(z) + 2(2.5s - 1)\kappa(z)
    \label{eq:cosmicmag}
,\end{equation}
where the second term on the right-hand side of the equation is the cosmic magnification contribution. Here $\kappa(z)$ is the line-of-sight integral of the lensing convergence to the galaxy redshift $z$; $s$ is the slope of the logarithmic cumulative number counts of our galaxy sample at magnitude limit $m_{\mathrm{lim}}$
\begin{equation}
    \left.s \equiv \frac{\partial \log _{10} N(<m)}{\partial m}\right|_{m=m_{\lim }}
    \label{eq:cdf}
.\end{equation}
The KiDS gold galaxy sample has an $r-$band magnitude limit of 25, but the completeness limit is around 24 \citep{wright2018kids+}. To properly estimate $s$, one needs a galaxy sample from a deeper survey that has a completeness magnitude of at least 25, as well as the same redshift distribution in each tomographic bin as the KiDS gold galaxy sample. In addition, the slopes for different tomographic bins should be different, but we do not have a good way to estimate them, so we take a simple method to estimate $s$. Given that the logarithmic cumulative number counts
of the galaxy sample is nearly linear with respect to $m$ \citep[see Fig. 6 of][]{wright2018kids+}\footnote{The paper cited here is a KV-450 paper, but this consideration applies equally to KiDS-1000 since the data depth is the same.}, we estimate $s$ by extrapolating the slope at the completeness magnitude (which is 24) to the magnitude limit, which is 25 \citep{2016MNRAS.455.3943H}. The resulting slope is 0.29. \revised{We also estimate the error of $s$ by assigning Poisson error in the count bins and measure an error on the slope. This yields an error of 0.001, which is subdominant.} We also try other magnitudes to extrapolate from and find that our best-fitting parameters only change marginally. We note that the $2\times2.5s\kappa(z)$ term accounts for flux magnification and $2\times\kappa(z)$ term accounts for angular dilution, so when $s=0.4$ both effects cancel out.

\begin{figure*}
    \centering
    \includegraphics[width=0.95\textwidth]{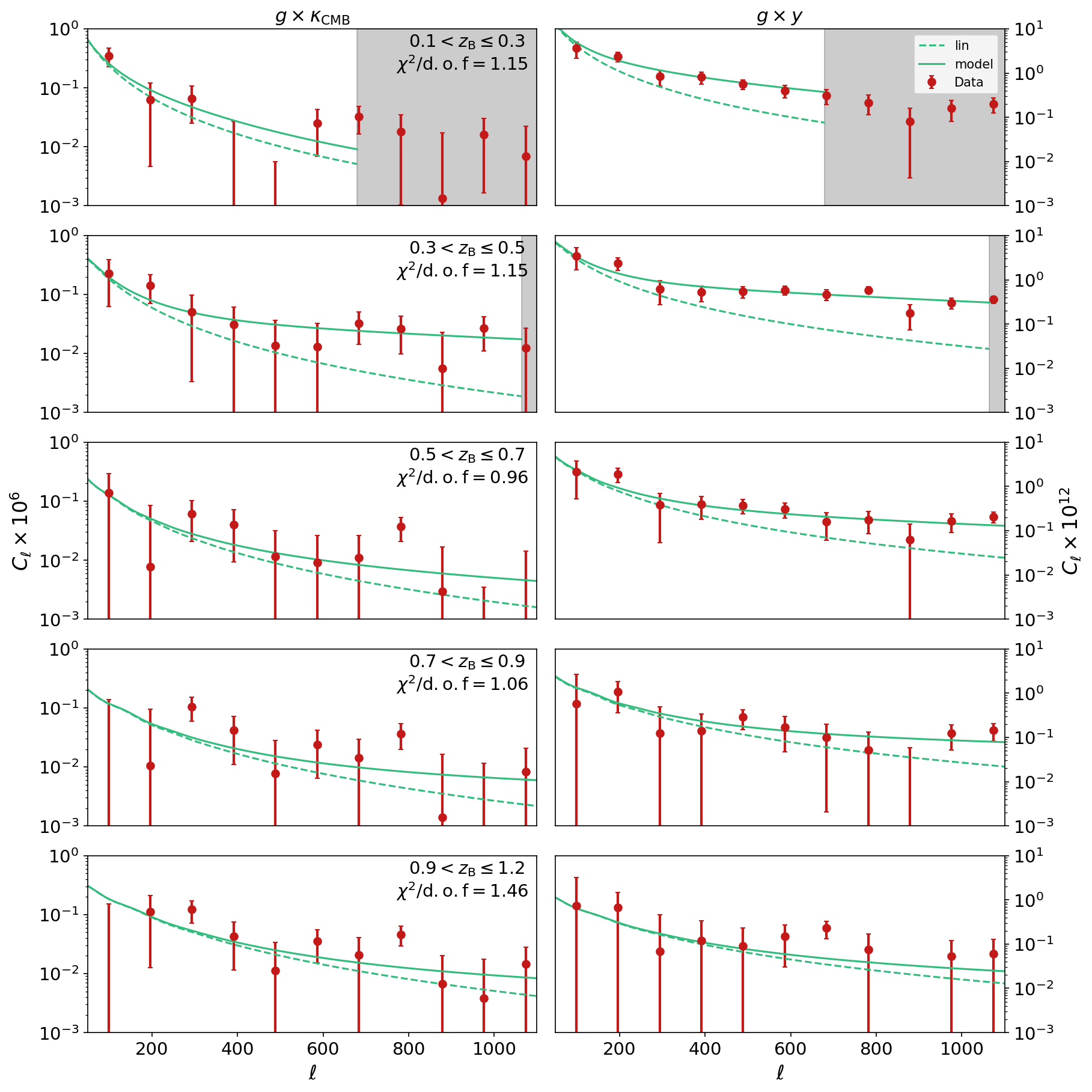}
    \caption{Measurements of $C^{\mathrm{g}\kappa}_{\ell}$ and $C^{\mathrm{g}y}_{\ell}$ (red dots with error bars) over-plotted with best-fit models(green lines). The non-linear model shown here is the halo model non-linear template. \revised{We also plot the linear part with dashed lines}. Each row shows results in each tomographic bin. Shaded regions are scales corresponding to $\ell$ scales within the threshold $k_{\mathrm{cut}} > 0.7\, \mathrm{Mpc}^{-1}$, which are not included the model-fit.}
    \label{fig:gygk_results}
\end{figure*}

\subsubsection{Uncertainty of the redshift distribution}

Uncertainties in the galaxy redshift distributions could affect galaxy cross-correlations. We estimate the uncertainties of the SOM redshift distributions using the same method as described in \citet{hildebr2020kids1000}, which gives uncertainties of the mean redshift at a level of $\sim 0.02$ in all 5 tomographic bins. To evaluate the impact of this uncertainty, we shift the fiducial redshift distribution $n_{\mathrm{g}}(z)$ in \eqref{eq:wg} by $\delta_z=\{-0.02, -0.01, 0, 0.01, 0.02\}$. We fit the model parameters with these shifted redshift distributions to see if this changes our results.

\section{Results}
\label{sect:results}

\begin{figure}
    \centering
    \includegraphics[width=\columnwidth]{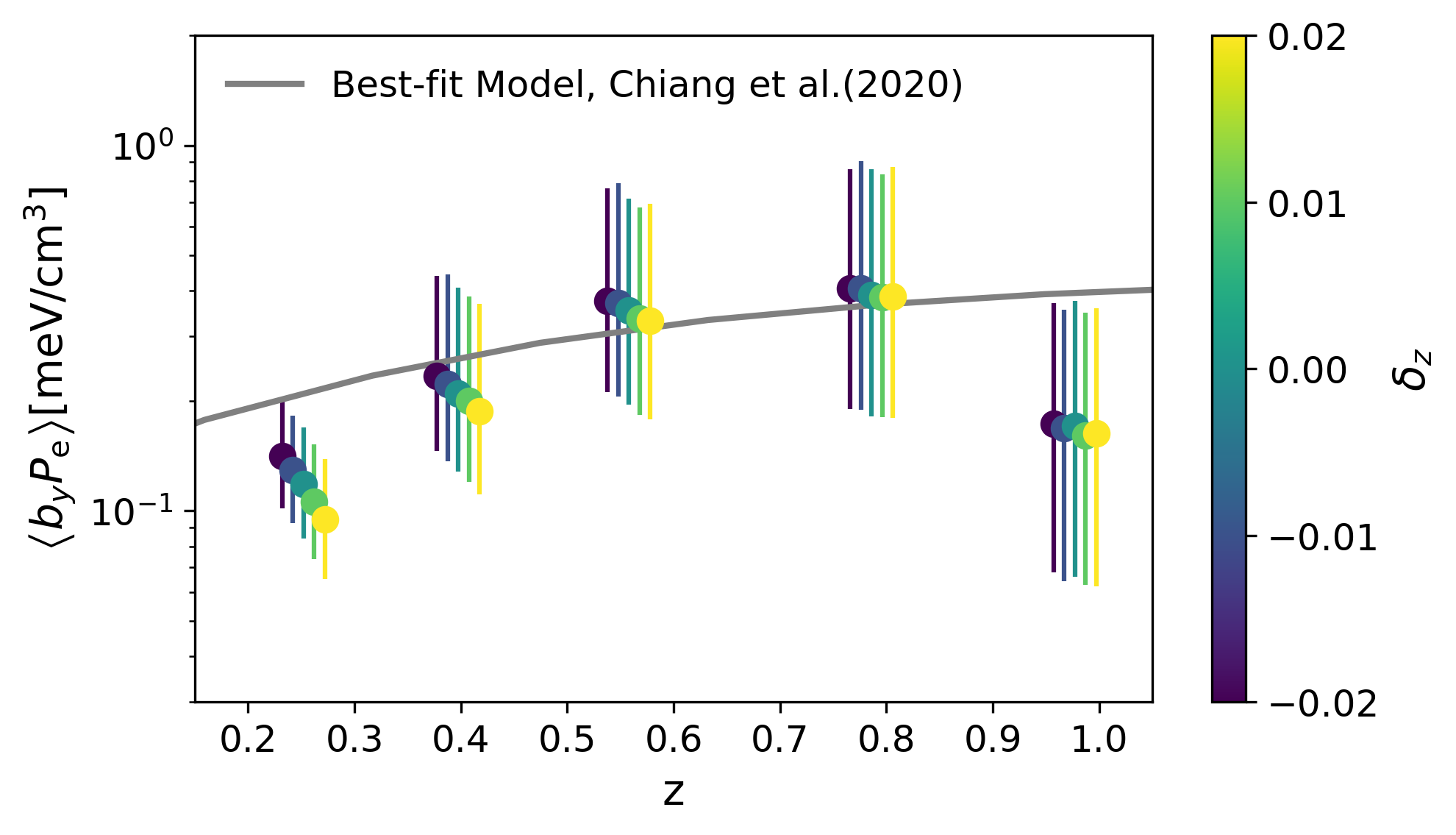}
    \caption{Best fit $\bpe$ with varying redshift distribution shifts. The non-linear model is the halo model non-linear template. The best-fit parameter values and errors are calculated as the modes and standard deviations of the Gaussian kernel density estimation (KDE) fits of marginalised posterior distribution.}
    \label{fig:gygk_deltaz}
\end{figure}

\begin{figure}
    \centering
    \includegraphics[width=\columnwidth]{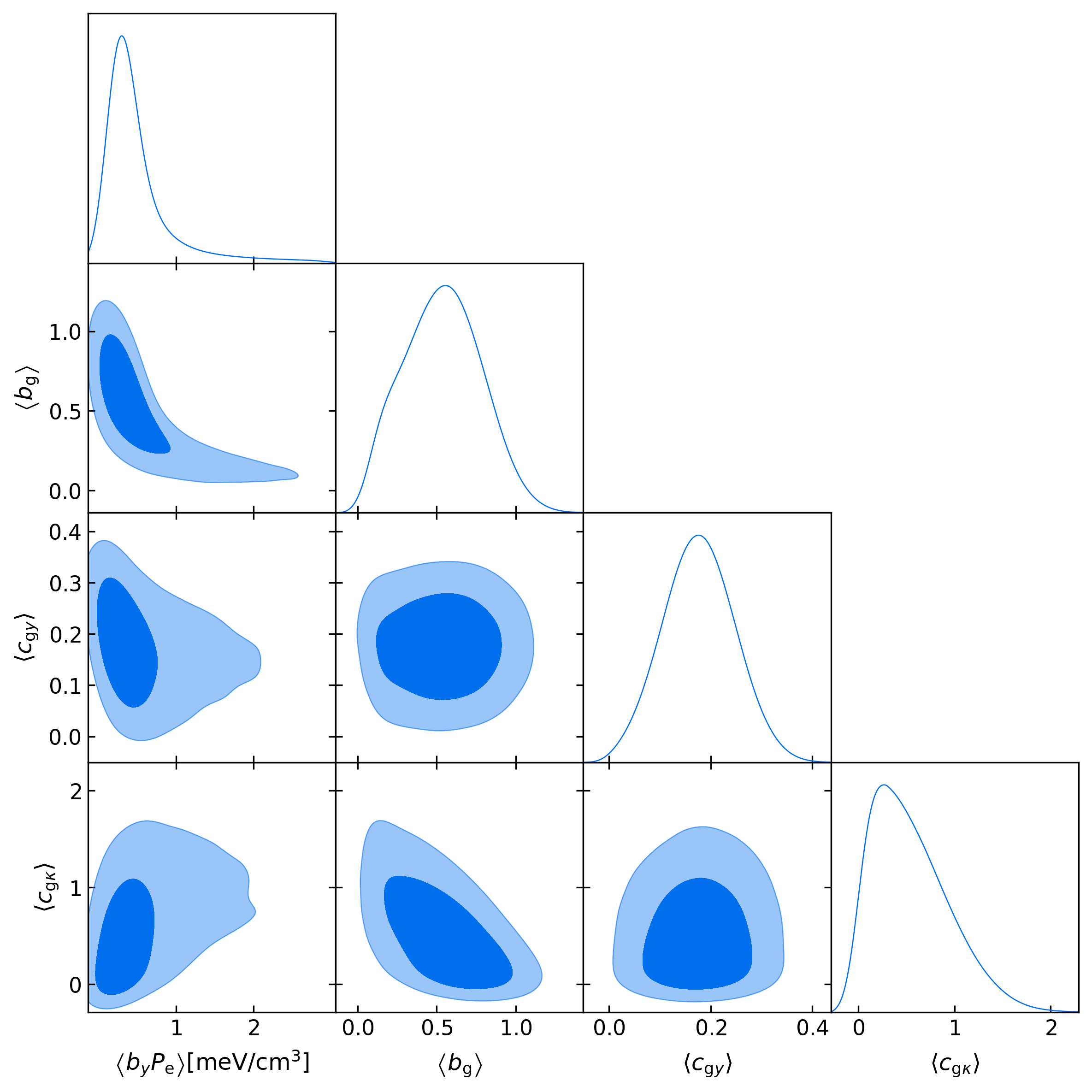}
    \caption{68\% and 95\% contours of the posterior distribution of $\left\{\bpe, \bg, \cgy, \cgk\right\}$ in the third redshift bin with the halo model non-linear template in the third redshift bin. The contours and 1-D posteriors have been smoothed for the purpose of presentation.}
    \label{fig:mcmc_post}
\end{figure}

\begin{figure*}
    \centering
    \includegraphics[width=\textwidth]{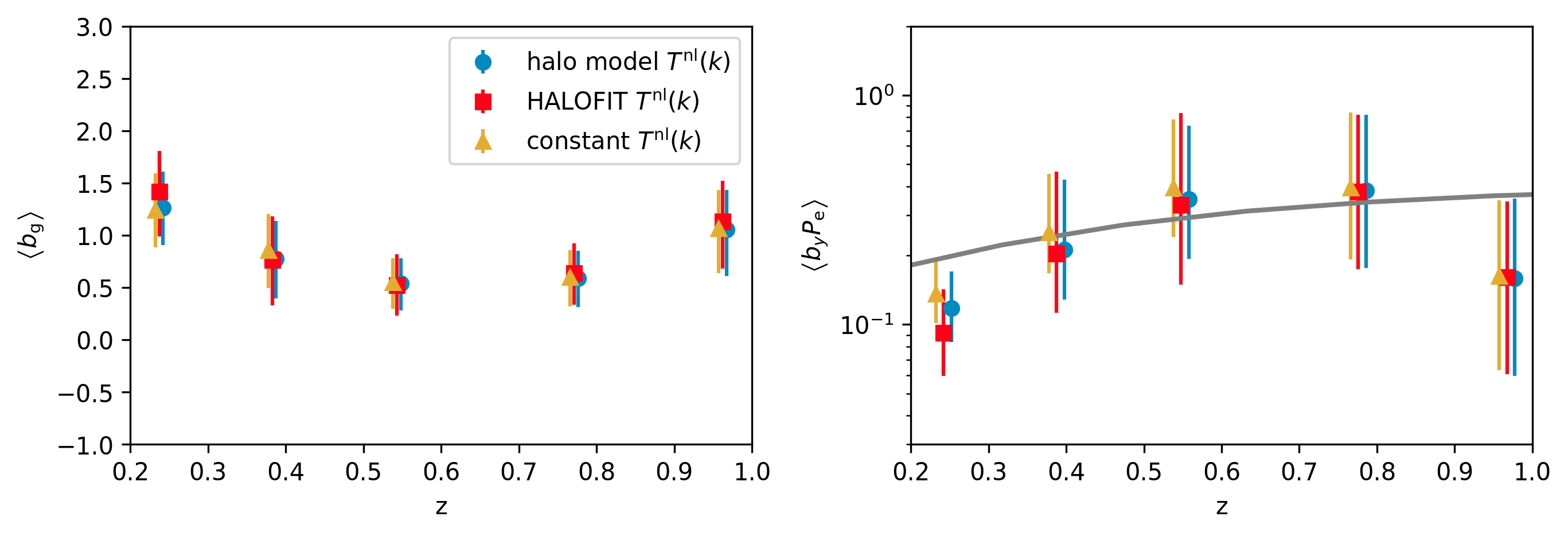}
    \caption{Constraints of $\bg$ and $\bpe$ in each tomographic bin. The best-fitting parameter values and errors are calculated as the modes and standard deviations of the Gaussian KDE fittings of marginalised posterior distributions. Dots with different colours are correspond to the different non-linear power spectrum models. The grey line shows the best-fit model from \citet{chiang2020cosmic}.}
    \label{fig:params_nlmodel}
\end{figure*}

\begin{figure*}
    \centering
    \includegraphics[width=1\textwidth]{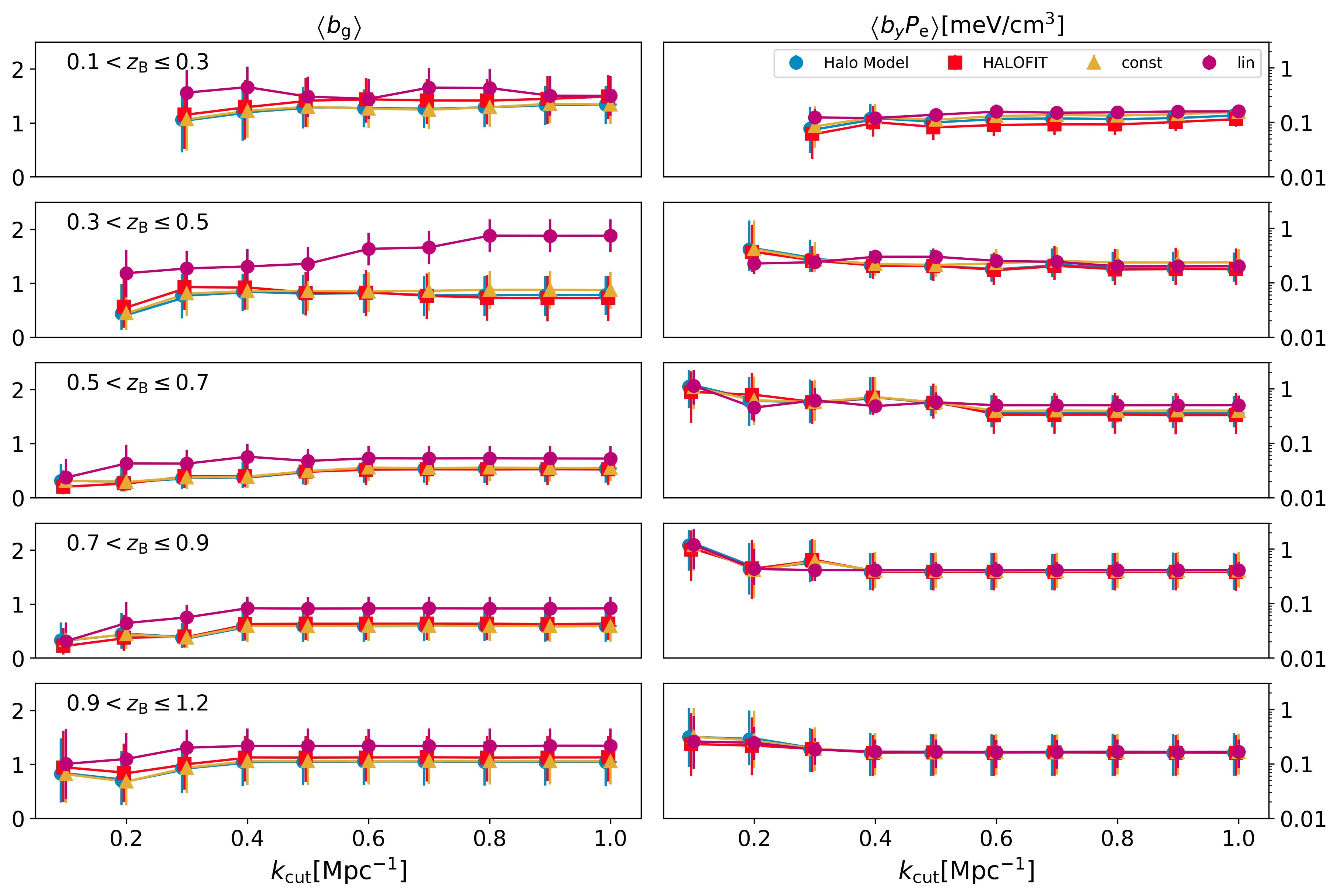}
    \caption{Constraints of $\bg$ and $\bpe$ for different non-linear matter power spectrum models with different scale cuts $k_\mathrm{cut}$. We also plot the fitting of $\bg$ and $\bpe$ with a pure linear model in purple (i.e. $\cgk$ and $\cgy$ are both fixed to be zero). We note that for low redshifts and low $k_\mathrm{cut}$, we do not have enough degrees-of-freedom so $\bg$ and $\bpe$ are not presented. In addition, for high redshift bins, large-scale cuts are beyond the high limit of $\ell$, so with these $k_\mathrm{cut}$ values, the constraints do not change.}
    \label{fig:params_kcut}
\end{figure*}

\begin{figure}
    \centering
    \includegraphics[width=1\columnwidth]{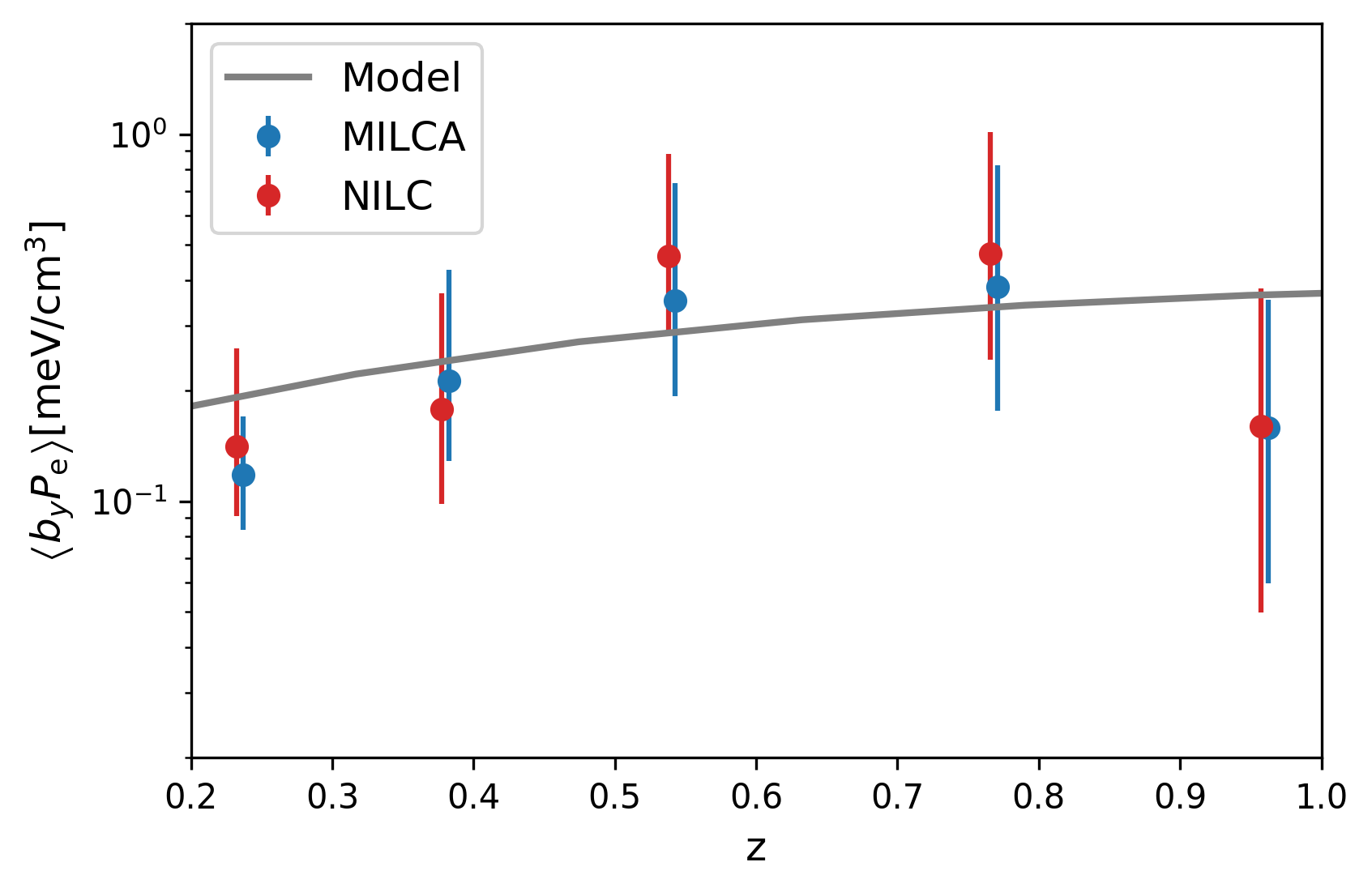}
    \caption{\revised{Constraints of $\bpe$ with \planck\xspace MILCA and NILC $y$ map.}}
    \label{fig:ymap_compare}
\end{figure}

We estimate the galaxy-tSZ and the galaxy-CMB lensing cross-correlations of KiDS galaxies as well as corresponding covariance matrices in each of the 5 tomographic bins with the methods described in Sect.  \ref{sect:measurements}. Figure \ref{fig:gygk_results} shows our measurements of $C^{\mathrm{g}\kappa}_{\ell}$ (left column) and $C^{\mathrm{g}y}_{\ell}$ (right column) bandpowers with red dots. The bandpowers are calculated as the mean $C_{\ell}$ in each $\ell$ bin and the error bars are given by the square root of the diagonals of covariance matrices. Each row corresponds to one of the tomographic bins. The measured cross-correlations are over-plotted with the best-fit fiducial model (with the halo model non-linear matter power spectrum) with green lines. Shaded regions are angular scales corresponding to $k_{\mathrm{cut}} > 0.7\, \mathrm{Mpc}^{-1}$, which are not included in the model-fit. \footnote{For each redshift bin, the $k$ threshold translates into the $\ell$ threshold via $\ell_{\mathrm{cut}}\equiv k_{\mathrm{cut}} \chi(z_{\mathrm{mean}})$, where $z_{\mathrm{mean}}$ denotes the mean redshift of each redshift bin. We note that this threshold is beyond the upper bound of $\ell$ for the last three redshift bins.} We set this threshold because we do not think our simple models will capture the details of the non-linear regime. We note that the red dots show our fiducial results with both CIB contamination and cosmic magnification corrected. In all the tomographic bins, CIB contributions are at a level of $\sim 1\%$ even with the most conservative level of $\alpha_{\mathrm{CIB}}$ (with $\alpha_{\mathrm{CIB}}=(2.3 + 6.6) \times 10^{-7}(\mathrm{MJy} / \mathrm{sr})^{-1}$), and can be neglected. We validate this claim by also fitting the raw, CIB contaminated $gy$ cross-correlation with our model and find no significant difference between the best-fit $\bpe$ values and the fiducial fitting. In order to evaluate the impact of cosmic magnification, we also fit a model without cosmic magnification correction given by Eq. \eqref{eq:cosmicmag}. Comparing with our fiducial results, we find that if the cosmic magnification is neglected, $\bg$ in all the redshift bins will be slightly under-estimated at a level of $\sim 1\%$. We also show the best-fit parameters from shifted redshift distributions in Fig. \ref{fig:gygk_deltaz}. Points with error bars in different colours correspond to different shifts of redshift distributions $\delta_z$. Our fiducial results have $\delta_z=0$. From the plot, we conclude that a redshift bias of $\delta_z\sim 0.02$ would only have a marginal effect on our results. It should be noted that the constraint in the first redshift bin gets mostly affected. This might be due to the fact that the redshift distribution of this bin is the narrowest, which makes it more sensitive to a redshift error.

An example of the fiducial MCMC posterior (that corresponds to the halo model non-linear power spectrum) distribution of $\{\bpe, \bg, \cgy, \cgk \}$ is shown in Fig. \ref{fig:mcmc_post}. We note that the linear biases that we are fitting are the normalisation of the two linear cross-correlations, namely $\{\bg\times\bpe, \bg\}$. Their posteriors are Gaussian in the linear region because they both linearly depend on the linear cross-correlations. The gas pressure bias $\bpe$ is then the ratio of two Gaussian parameters, so its posterior distribution is asymmetric, as shown in Fig. \ref{fig:mcmc_post}. A summary of the fiducial fitting results of $\bg$ and $\bpe$ is given in Table \ref{table:fitinfo}. The best-fit parameter values and errors are calculated as the modes and standard deviations of the Gaussian kernel density estimation (KDE) fittings of the marginalised posterior distributions. We evaluate the constraining power of both linear bias parameters with the method given by \citet{Asgari_2021}. That is, we calculate the values of the marginalised posterior at both extremes of the prior distribution, and compare them with 0.135, the ratio between the peak of a Gaussian distribution and the height of the $2\sigma$ confidence level. If the posterior at the extreme is higher than 0.135, then the parameter boundary is not well constrained. We find that the lower bound of $\bpe$ of the fifth redshift bin does not meet this criterion (this can also be seen from the fact that the $2\sigma$ lower bound of $\bpe$ in the fifth redshift bin is below zero). However, it should be noted that the lower extremes of the bias parameters are physical limits, which could not be extended. For the parameters that are not bounded, KDE might give inaccurate results. To test that, we calculate the best-fit $\bpe$ in the fifth redshift bin without smoothing the marginalised posterior with a KDE kernel. This changes the value from 0.16 to 0.134. The difference is well below the constraining error.

In order to evaluate the goodness-of-fit, we calculate the $\chi^2$ in each tomographic bin, and then calculate the corresponding probability-to-exceed (PTE) given the degree-of-freedom. \citet{heymans2020kids1000} adopts the criterion PTE>0.001 (corresponding to a $\sim3\sigma$ deviation) to be acceptable. We find our fittings of all the tomographic bins with all the three non-linear models meet this criterion. The pure linear model does not fit well, especially in low redshift bins. Thus we conclude that all of our non-linear models fit well with our data. 

We show the redshift dependence of $\bg$ and $\bpe$ in Fig. \ref{fig:params_nlmodel}. The dots with different colours are the constraints from our results with three non-linear power spectrum models, marginalised over the non-linear parameters $\cgy$ and $\cgk$. The plots show that the constraints on $\bg$ and $\bpe$ are consistent with different non-linear power spectrum models, indicating that our measurements are insensitive to the details of the non-linear cross-correlations. To further verify this argument, we repeat the model fitting with different scale cuts $k_{\mathrm{cut}}$ (modes of $C_{\ell}$ with a scale smaller than $k_{\mathrm{cut}}$ are removed from the model fitting procedure) and plot the best-fit values of $\bg$ and $\bpe$ as a function of $k_{\mathrm{cut}}$ in Fig. \ref{fig:params_kcut}. The plot shows that the constraints do not change significantly as different scales are removed, so we conclude that our constraints are robust to non-linear details. To highlight the importance of transition region error, we also fit the halo model and the constant non-linear model without correction with $R(k)$ defined in Eq. \eqref{eq:rkdef}. We find that this correction changes the best-fit parameter value by a few percent, worst in the lower redshift bin (about 10\%), but the differences are below the 1$\sigma$ level. This is because the {\it{Planck}} beam size ensures that the data are noisy at these scales. However, future studies with higher resolution should be sensitive to these systematics. We also acknowledge that different non-linear models affect $\bpe$ in the lowest redshift at a level of $0.5-1\sigma$ because {\it{Planck}} beam does not smooth out the non-linear details as completely as high redshift bins. In Fig.~\ref{fig:params_kcut}, we also plot the fitting of $\bg$ and $\bpe$ with a pure linear model in purple (i.e. $\cgk$ and $\cgy$ are both fixed to be zero). We find that the pure linear model gives $\bg$ values that are higher than non-linear models on all scales, and have the tendency to merge with the fiducial fitting with low $k_{\mathrm{cut}}$. The gas pressure bias $\bpe$ is the ratio between linear amplitudes of $g\kappa$ and $gy$ cross-correlations, so it could be close to the real value even if the linear model gives biased amplitudes. This result highlights the necessity to include some form of non-linear model in the fitting.

In our analysis, we use the whole KiDS gold lensing galaxy sample, in which there are many blue galaxies that are distributed out to a large distance from cluster centres \citep{Croton_2007}. The best-fit $\bg$ values in all the redshift bins are consistent with one, which suggests that our galaxy sample is a good tracer of the dark matter distribution. This should be contrasted with, for example,
luminous red galaxies (LRG), which are strongly biased tracers of mass \citep{2005ApJ...621...22Z} because LRGs are known to be clustered around halo centres.

\revised{We test the robustness of our model fitting to different Compton $y$ reconstruction method by replacing the \planck\xspace MILCA $y$ map with the NILC $y$ map. The constraints on $\bpe$ is shown in Fig. \ref{fig:ymap_compare}, which indicates a consistency between two $y$ maps. Thus we conclude that our measurement is not sensitive to different $y$ reconstruction methods.}

\begin{figure}
    \centering
    \includegraphics[width=\columnwidth]{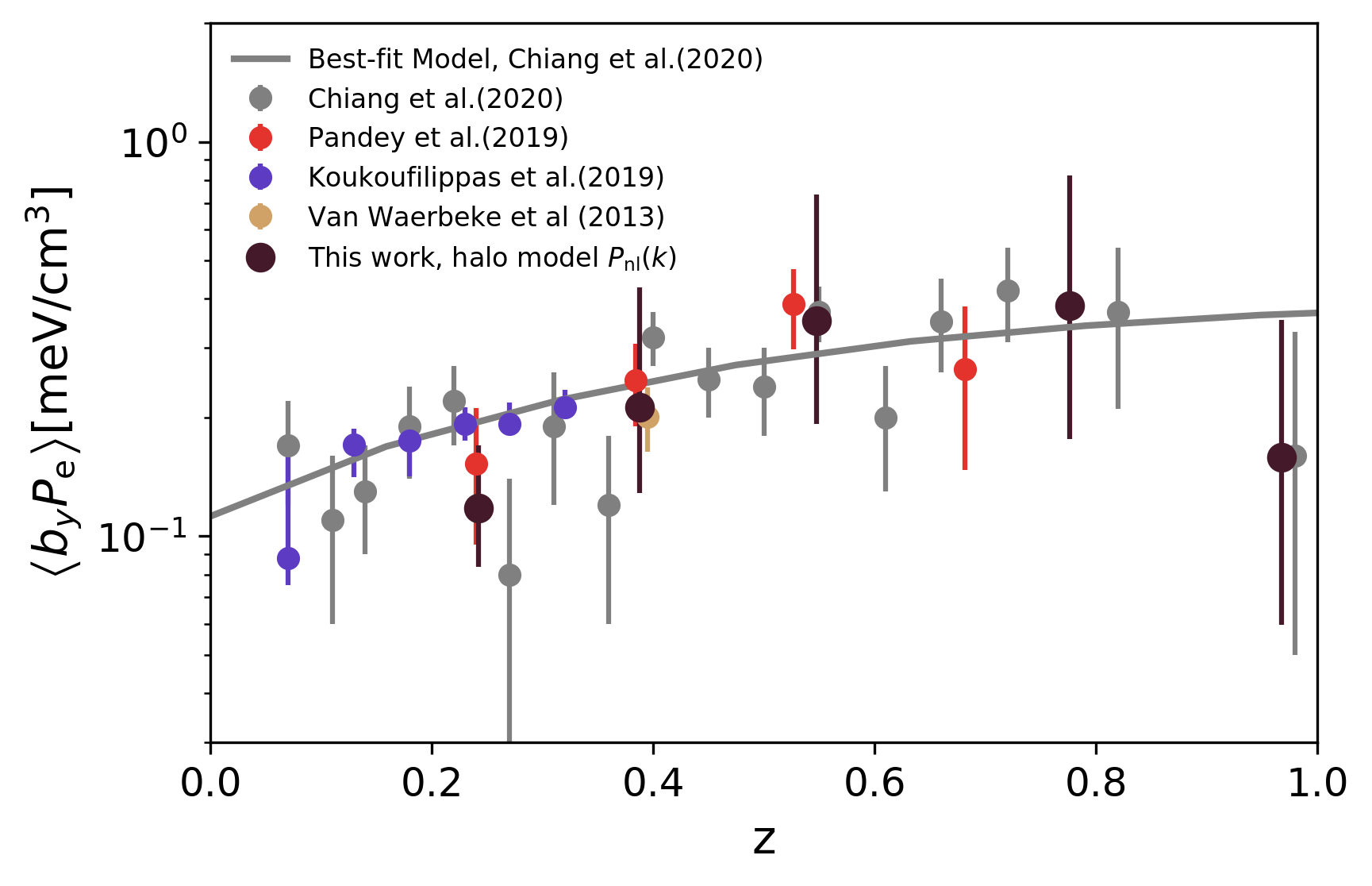}
    \caption{Constraints of $\bpe$ in each tomographic bin. Our results with the halo model non-linear matter power spectrum is presented as black dots with error bars. The best-fit parameter values and errors are calculated as the modes and standard deviations of the Gaussian KDE fittings of marginalised posterior distributions. Results from previous studies are also plotted as well as the best-fit $\bpe$ model given by \citet{chiang2020cosmic}.}
    \label{fig:bpe_fit}
\end{figure}

Fig.~\ref{fig:bpe_fit} compares our constraints on $\bpe$ to previous studies. These studies relied on the cross-correlation of data from different surveys with {\textit{Planck}} tSZ data. \citet{van2014detection} (orange dot) uses the lensing data from the RCSLenS sample; \citet{koukoufilippas2020tomographic} (purple dots) uses the 2MPZ and WISE$\times$SuperCosmos samples; \citet{pandey2019constraints} (red dots) uses the DES sample; \citet{chiang2020cosmic} uses the galaxy samples from SDSS, BOSS for low redshifts, and QSO samples from SDSS, BOSS and eBOSS for high redshifts. All these studies used the galaxy auto-correlation to measure $\bg$, except \citet{van2014detection} who used the mass distribution measured from weak gravitational lensing. In our approach, we use the galaxy-CMB lensing cross-correlation to constrain $\bg$. It is remarkable that these measurements, from very different surveys and with very different estimators give consistent results for $\left\langle b_{y} P_{\mathrm{e}}\right\rangle$. However, it should be noted that although our measurement goes to higher redshift, our constraining power is significantly weaker than previous studies in the same redshift range. This is because 1) the limitation of the KiDS footprint makes it less sensitive to linear scales; 2) the CMB lensing map is noisy. With future sky surveys having wider sky coverage and CMB surveys having lower noise levels, these two drawbacks can be improved. The grey line in Fig.~\ref{fig:bpe_fit} is the best-fit redshift dependence of the tSZ halo model given by \citet{chiang2020cosmic}. Although this work does not constrain that model, it is introduced and discussed in Appendix~\ref{app:forecast} for forecasting future studies. With an agreement with these previous results as well as the halo model prediction, our tomographic measurement provides insights into the thermal history of the LSS. 

\begin{table*}
\centering

\begin{tabular}{llllll}
\toprule
$Z_{\mathrm{bin}}$ & $z_{\mathrm{mean}}$ &                   $b_{\mathrm{g}}$ & $\left\langle bP_{\mathrm{e}}\right\rangle$[meV/cm$^3$] & $\chi^2/{d.o.f}$ &   PTE \\
\midrule
 0.1$ < Z_\mathrm{B} \leq $0.3 &                0.23 &  1.26$^{+0.34}_{-0.36}$ &                       0.12$^{+0.05}_{-0.03}$ &           1.15 &  0.22 \\
 0.3$ < Z_\mathrm{B} \leq $0.5 &                0.38 &  0.78$^{+0.36}_{-0.38}$ &                       0.21$^{+0.22}_{-0.08}$ &           1.15 &  0.39 \\
 0.5$ < Z_\mathrm{B} \leq $0.7 &                0.54 &  0.54$^{+0.24}_{-0.26}$ &                       0.35$^{+0.38}_{-0.16}$ &           0.96 &  0.53 \\
 0.7$ < Z_\mathrm{B} \leq $0.9 &                0.77 &  0.59$^{+0.26}_{-0.28}$ &                       0.38$^{+0.44}_{-0.21}$ &           1.06 &   0.4 \\
 0.9$ < Z_\mathrm{B} \leq $1.2 &                0.96 &  1.05$^{+0.38}_{-0.45}$ &                         0.16$^{+0.2}_{-0.1}$ &           1.46 &  0.09 \\
\bottomrule
\end{tabular}

\caption{Best-fitting linear bias parameters from each tomographic bin. The results correspond to the non-linear power spectrum model being the default halo model. The best-fit parameter values and errors are calculated as the modes and standard deviations of the Gaussian KDE fit of marginalised posterior distributions. PTE stands for the probability-to-exceed of the corresponding reduced $\chi^2$ value.}
\label{table:fitinfo}

\end{table*}

The linear bias assumption might break down on small scales where baryonic effects become significant. However, in our analysis, the {\it{Planck}} beam makes our measurements insensitive to these effects. We leave as a future work the generalisation of our results to the small scales when data with higher resolution are available. Such a situation can be handled with a more sophisticated model, for example that from \citet{mead2020hydrodynamical}.

In our study, our approach consists of using CMB lensing as a way to constrain $\bg$ of a galaxy sample. One can independently measure $\bg$ using weak gravitational lensing as a replacement for CMB lensing. Appendix \ref{sec:ggl} shows the results when CMB lensing is replaced by the KiDS weak lensing signal. We find that the constraints on $\left\langle b_{y} P_{\mathrm{e}}\right\rangle$ are consistent with CMB lensing. This result validates our approach and highlights the fact the CMB lensing and galaxy lensing can be used independently to calibrate the mass of a galaxy distribution using a very different source redshift screen. Fig.~\ref{fig:param_ggl} also shows that the highest redshift bin is significantly noisier when galaxy lensing is used compared to CMB lensing, while galaxy lensing provides higher signal-to-noise for lower redshifts. This is a direct consequence of the very different source redshift between CMB and galaxy lensing, and it illustrates the fact that lensing signal-to-noise decreases dramatically when the lenses are close to the sources, as expected.

\section{Discussion and conclusion}
\label{sect:discussions}

In this work we use the galaxy sample from the fourth KiDS Data Release, the {\textit{Planck}} $y$ map and {\textit{Planck}} CMB lensing map to probe the redshift dependence of galaxy bias of KiDS galaxies and gas pressure bias from the galaxy$\times$tSZ and the galaxy$\times\kappa_{\mathrm{CMB}}$ cross-correlations. We assume that, in the linear region, both tSZ $y$ parameter and galaxy overdensity are proportional to the underlying mass fluctuation, with the proportionality parametrised by galaxy bias $\bg$ and gas pressure bias $\bpe$, which is consistent with our measurement being restricted to large angular scales. To account for the non-linear effects, we also model the non-linear power spectra of $g y$ and $g \kappa$ cross-correlations as rescaled non-linear templates. We tried three kinds of non-linear templates: halo model, \halofit, and constant, all of which yield consistent constraints of $\bg$ and $\bpe$, indicating that our measurements are not yet sensitive to the non-linear details. However, with an additional inconsistent constraint with a purely linear model, we emphasise the necessity to consider non-linear cross-correlations. $\bg$ and $\bpe$ are constrained for galaxies from five tomographic bins within $z\lesssim 1$, which counts amongst the furthest distance probed from this kind of analysis. The reduced $\chi^2$ of the best-fit parameter values indicate that our model fits the data well. 

The best-fitting galaxy bias is close to 1, which indicates that the KiDS galaxy sample is an unbiased tracer of the underlying mass distribution. In previous works \citep[for example][]{koukoufilippas2020tomographic, chiang2020cosmic,pandey2019constraints} the authors used galaxy auto spectra to constrain galaxy bias, which is subject to modelling uncertainties and auto-correlated noise. Our approach avoids this problem by using the CMB lensing to calibrate the mass from the galaxy distribution. In Appendix~\ref{sec:ggl}, we will show that our results are unchanged when we replace CMB lensing by galaxy lensing from KiDS. 

Fig. \ref{fig:bpe_fit} shows our constraints on $\bpe$ in each of the tomographic bins as well as the measurements from previous studies \citep{van2014detection,pandey2019constraints, koukoufilippas2020tomographic, chiang2020cosmic}. Our result agrees well with them. We also compare our result with predictions of the halo model \citep{chiang2020cosmic} and find good agreement. Our tomographic measurement of $\bpe$ confirms the evolution of biased thermal energy in halos into the high redshift regime. In addition, the gas bias $b_{y}$, estimated to be $\sim 3.5$ \citep{chiang2020cosmic}, parametrises the link between gas and dark matter halo; the mean electron pressure $\left\langle P_{\mathrm{e}} \right\rangle=\left\langle n_{\mathrm{e}} \right\rangle k_B\overline{T}_e$ is associated with the thermal dynamic property of electrons. Based on CMB constraints, the average electron number density is $\left\langle n_{\mathrm{e}} \right\rangle \sim 0.25\,\mathrm{m}^{-3}$ \citep{Hinshaw_2013}. Taking these values into account, the mean electron temperature $\overline{T}_e$ is at a level of $\overline{T}_e\sim 10^6$ K, which is consistent with the estimated temperature of `missing baryons'\citep{cen1999baryons}. This means that if the tSZ signal were entirely from intergalactic gas, it could account for all the missing baryons within the temperature range $10^5-10^7$ K \citep{Bregman_2007}. To confirm this, we need a halo model for diffuse baryons that can properly describe the spatial distribution of gas within dark matter halos, which we leave to future work. Our study consolidates our understanding of intergalactic gas at high redshift, which, combined with future tomographic measurements on $\bpe$, will improve our understanding on the thermal history of the Universe, as well as the evolution of links between gas and dark matter halos. 

The uncertainty we find for $\bpe$ is larger than previous studies because KiDS has a smaller sky coverage compared with those surveys, which makes it less sensitive to the linear regime where the majority of constraining power lies. Uncertainties of $\bg$ and $\bpe$ are both dominated by sample variance on the linear scales. Future sky surveys such as the Rubin Observatory Legacy Survey of Space and Time (LSST) \citep{abell2009lsst} and the \euclid\, survey \citep{laureijs2010euclid} will cover a larger fraction of the sky, making it possible to yield tighter constrain on linear biases. In addition, future CMB-S4 and Simons Observatory-like experiments will provide CMB lensing and $y$ maps with lower noise levels \citep{Hadzhiyska_2019} and with higher angular resolution, which will improve both the constraining capacity of galaxy bias and sensitivity to small-scale physics. With these improvements, one could model the non-linear cross-correlation with a more sophisticated model, such as the full halo model with a GNFW profile \citep{Arnaud_2010} for the tSZ and the HOD model for the galaxy distribution. We make a forecast on $g\kappa$ and $g y$ cross-correlations with such a model with hypothetical LSST, \euclid $\,$, and CMB-S4 surveys in Appendix \ref{app:forecast}, which yields a tight constraint on $\bpe$. Our forecast highlights the validity of multi-tracer analysis for future sky surveys.

We carefully evaluate the systematics in our data that could cause bias in our model fitting. The main systematics considered in this study are the cosmic magnification in galaxy overdensity measurements, CIB contamination in tSZ map and uncertainties in the redshift distributions. Though all of these systematics are not significant in our measurement due to low signal-to-noise in our data, they will become significant for future surveys with large sky coverage. In principle, the first two systematics affect high redshifts more significantly, so future studies with a deeper redshift reach must carefully take these into account. It should also be noted that the parameter constraints in this study are quoted under the assumption of fixed cosmological parameters from {\textit{Planck}} \citep{aghanim2018planck}. The amplitude of angular cross-correlations is closely related to galaxy and gas biases as well as $\sigma_8$ \revised{and possible reconstruction bias in the CMB lensing map}. So these parameters are strongly degenerate, \revisednew{and this is the fundamental limit of this analysis}. We could robustly constrain cosmological parameters as well as galaxy and gas biases by combining more cross-correlation measurements, like galaxy clustering and cosmic shear. Once again, we leave this to future studies.

This work shows the potential to study LSS by combining different cross-correlations measurements. cross-correlation is known to be immune to auto-correlated noise. A combination of different cross-correlations can break the degeneracies between model parameters. Specifically, in our fiducial measurements, we do not use cosmic shear, which is affected by intrinsic alignments and shape miscalibration. Instead, we use CMB lensing as a non-biased tracer of LSS to independently constrain $\bg$. We provide a sanity check in Appendix \ref{sec:ggl} by replacing the CMB lensing map with the KiDS shear map and perform the same analysis, which gives consistent results for all tomographic redshift bins; this validates our fiducial method. However, the results from galaxy lensing at high redshift are noisier than our fiducial results, which indicates the advantage of using CMB lensing as a proxy for the mass distribution, especially at high redshift. Future work could combine CMB lensing and galaxy lensing as independent mass tracers, which could yield tighter constraints on LSS properties. Future surveys will also provide denser galaxy samples in wider ranges and deeper reaches in the sky, as well as cleaner CMB lensing and $y$ maps, which will make this method more promising for multi-tracer cosmology.

\begin{acknowledgements}
We thank Prof. Peter Schneider, Prof. Cheng Li, Prof. Houjun Mo, and the anonymous referee for fruitful discussions. 
ZY and LVW acknowledge support by the University of British Columbia, Canada’s NSERC, and CIFAR. 
We acknowledge support from the European Research Council under grant numbers 647112 (TT, MA, CH, AJM) and 770935 (AHW, HHi).
DA acknowledges support from the Beecroft Trust, and from the Science and Technology Facilities Council through an Ernest Rutherford Fellowship, grant reference ST/P004474. 
MB is supported by the Polish National Science Center through grants no. 2020/38/E/ST9/00395, 2018/30/E/ST9/00698 and 2018/31/G/ST9/03388, and by the Polish Ministry of Science and Higher Education through grant DIR/WK/2018/12.
CH acknowledges support from the Max Planck Society and the Alexander von Humboldt Foundation in the framework of the Max Planck-Humboldt Research Award endowed by the Federal Ministry of Education and Research. 
HHi is supported by a Heisenberg grant of the Deutsche Forschungsgemeinschaft (Hi 1495/5-1). 
NK is funded by the Science and Technology Facilities Council (STFC).
KK acknowledges support from the Royal Society and Imperial College
HYS acknowledges the support from NSFC of China under grant 11973070, the Shanghai Committee of Science and Technology grant No.19ZR1466600 and Key Research Program of Frontier Sciences, CAS, Grant No. ZDBS-LY-7013.\par 
\\
{{\it Author contributions:} All authors contributed to the development and writing of this paper. The authorship list is given in three groups: the lead authors (ZY \& LW) followed by two alphabetical groups. The first alphabetical group includes those who are key contributors to both the scientific analysis and the data products. The second group covers those who have either made a significant contribution to the data products, or to the scientific analysis. }
\end{acknowledgements}

\bibliographystyle{aa}
\bibliography{main}

\begin{appendix}
\section{Alternative method: Galaxy-galaxy lensing cross-correlations}
\label{sec:ggl}

In this appendix, we measure the cross-correlation between galaxy overdensity and lensing-induced galaxy shear caused by as an alternate to CMB lensing. This measurement serves as a sanity check of our fiducial result. We replace the CMB lensing map with the galaxy shear map generated with \revisednew{all the galaxies in the fifth} tomographic bin of the gold KiDS lensing shear catalogue and measure the cross-correlation between galaxy overdensity in each tomographic bin and galaxy lensing convergence $\kappa_{\mathrm{gl}}$. The kernel of $\kappa_{\mathrm{gl}}$ is given by:
\begin{equation}
    W^{\kappa_{\mathrm{gl}}}(\chi)=\frac{3 \Omega_{\mathrm{m}} H_{0}^{2}}{2 ac^{2}} g(\chi)
,\end{equation}
with
\begin{equation}
g(\chi)=\int_{\chi}^{\chi_{\mathrm{H}}} \mathrm{d} \chi^{\prime} n_{\mathrm{g}}\left(z(\chi^{\prime})\right) \chi\frac{\chi^{\prime}-\chi}{\chi^{\prime}}
,\end{equation}
where $\chi_{\mathrm{H}}$ is the comoving distance to the horizon. For the non-linear region, we take the rescaled halo model described in Sect. \ref{sect:model}.

The KiDS lensing shear catalogue \citep{giblin2020kids1000} provides ellipticities ($e_1$, $e_2$) of each galaxy. From them we construct a map triplet of ellipticities (0, $-e_1$, $e_2$) \citep{Harnois_D_raps_2017} as an analogue to CMB temperature-polarisation map triplet (T, Q, U). The minus sign on $e_1$ is due to the different convention of positive $x$ direction between lensing experiments and CMB polarisation experiments. We measure the cross-correlation between the KiDS galaxy overdensity maps and shear map triplet with \textsc{polspice}. The output `$\Delta_\mathrm{g} E$' mode (similar to the `TE' mode of CMB) is the cross-correlation between galaxy overdensity and galaxy-galaxy lensing convergence $g\kappa_{\mathrm{gl}}$ cross-correlation that we want. The `$\Delta_\mathrm{g} B$' mode (similar to the `TB' mode of CMB) should be zero, and we take it as a null test.

\begin{figure*}
    \centering
    \includegraphics[width=0.8\textwidth]{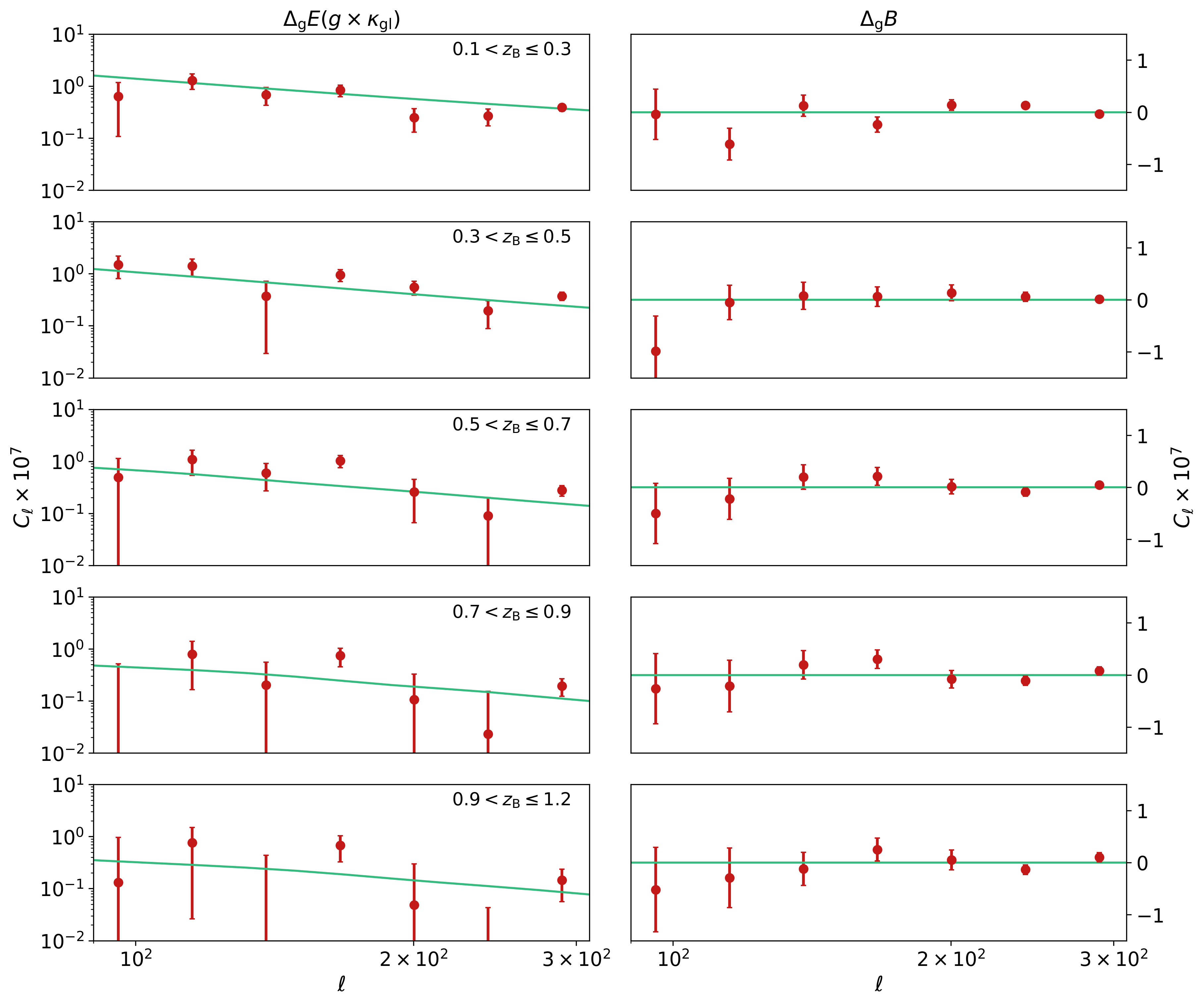}
    \caption{Galaxy-shear cross-correlation. The left column shows the `$\Delta_\mathrm{g} E$' mode, which corresponds to $g\kappa_{\mathrm{gl}}$ cross-correlation and the right column shows the `$\Delta_\mathrm{g} B$' mode as a null test. Green lines are the best-fit models. We find that the `$\Delta_\mathrm{g} B$' mode is consistent with zero.}
    \label{fig:ggl}
\end{figure*}

The covariance matrices are calculated the same way as our fiducial measurement. The cosmic magnification is also corrected. For scales with $k>0.3 \, h \mathrm{Mpc}^{-1}$, non-linear galaxy bias becomes significant \citep{heymans2020kids1000}. Besides, we want to match the physical scale of the cross-correlations between $g\kappa_{\mathrm{gl}}$ and $g\kappa_{\mathrm{CMB}}$. \revised{Intrinsic alignment is negligible in galaxy-galaxy lensing measurements \citep{Blazek_2012}.} Take these factors into account; we only use angular scales $100<\ell<200$ for the lowest redshift bin and $100<\ell<300$ for the rest. The measurements and best-fit models are plotted in Fig. \ref{fig:ggl}. The `$\Delta_{\mathrm{g}} B$' mode is consistent with zero in all redshift bins. The $g\kappa_{\mathrm{gl}}$ cross-correlation measured at high redshift are very noisy because for those bins the shear maps contain many galaxies that are in front of lenses, which only contributes to random noise from their intrinsic alignments as well as their cosmic shear from the foreground.

\begin{figure}
    \centering
    \includegraphics[width=\columnwidth]{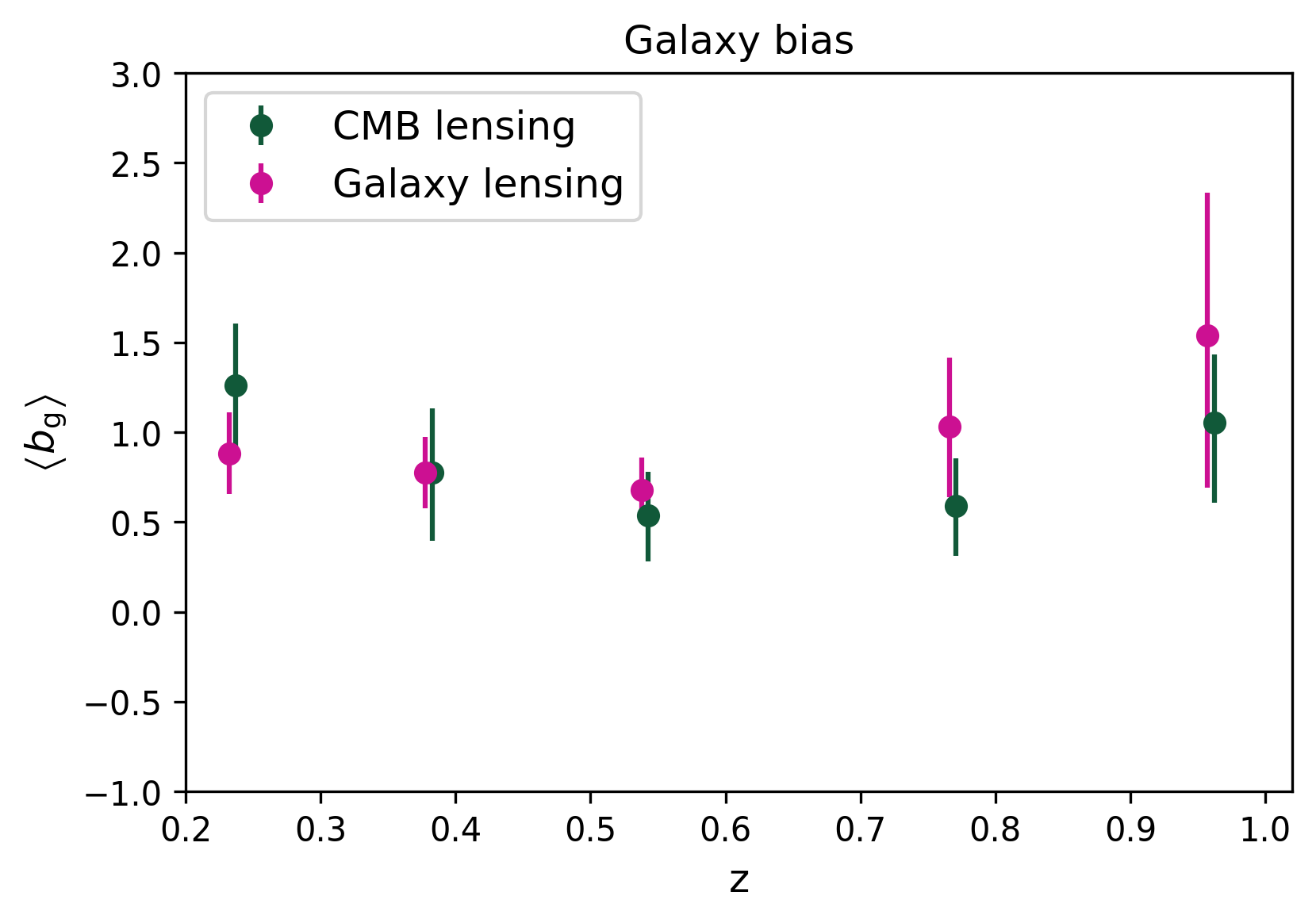}
    \includegraphics[width=\columnwidth]{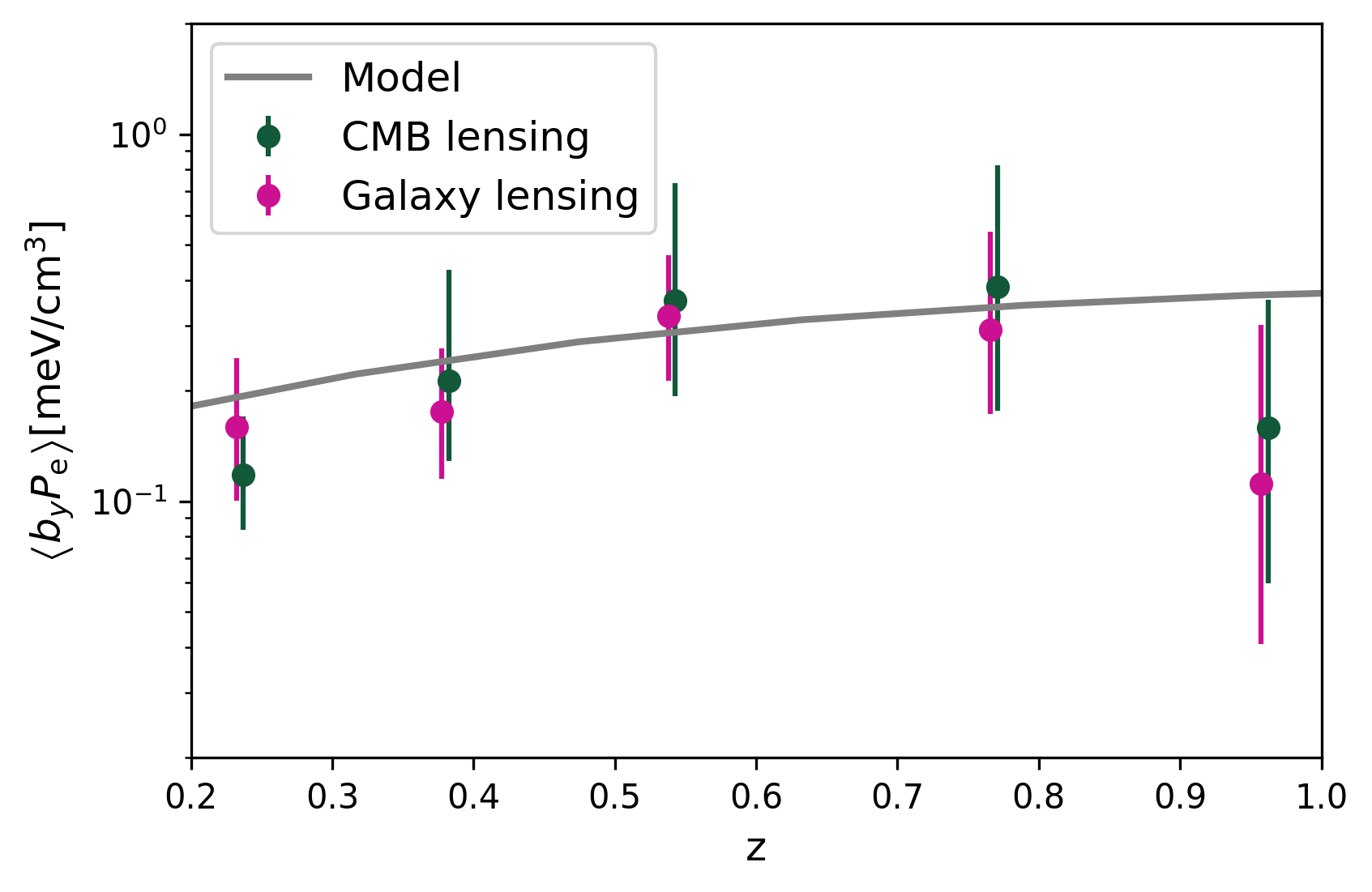}
    \caption{Constraints of model parameters with the galaxy-galaxy lensing and the galaxy-tSZ cross-correlations. Upper panel shows the constraints on $\bg$ and the lower panel shows $\bpe$ overplotted with halo model results from \citet{chiang2020cosmic}.}
    \label{fig:param_ggl}
\end{figure}

We present the constraints of our model parameters with the galaxy-galaxy lensing cross-correlation in Fig. \ref{fig:param_ggl}. The constraints on both $\bg$ and $\bpe$ are consistent. The errors of CMB lensing and galaxy lensing are comparable in the low redshift bins, which is expected if both mass proxies are cut at a comparable scale where the ellipticity noise is subdominant, and the error is mainly driven by sampling variance. This is a proof that CMB lensing and galaxy lensing are not only consistent (which is impressive given $\sim$12 billion years of separation between the two source populations), but also have comparable signal-to-noise in the sampling variance dominated regime. The uncertainty of the $\bg$ in the last two redshift bins is very high due to low signal-to-noise in these bins because in these bins most of the source galaxies are actually in front of lens galaxies. Thus they only contribute noise. This indicates that CMB lensing out-performs galaxy lensing when cross-correlated with high-redshift galaxies. This sanity check not only proves the reliability of our fiducial results but also shows the advantages of CMB lensing experiments over galaxy lensing with the KiDS data. For future studies, a combination of these measurements will yield even better constraint on models of interest.

\section{Forecasting the constraining power of full halo model for future sky surveys}
\label{app:forecast}
In this section, we forecast the constraining power on the full halo model \citep{COORAY_2002,Seljak_2000} parameters from $g\kappa$ and $g y$ cross-correlations measured from future sky surveys. The galaxy catalogue is assumed to be taken from a LSST/{\it{Euclid}}-like survey; the CMB lensing and tSZ data are from the CMB-S4 \citep{abazajian2016cmbs4} experiment. LSST and {\it{Euclid}} will cover a much wider area of the sky than KiDS, which makes them sensitive to the linear region as well as improves the signal-to-noise. CMB-S4 will achieve a higher angular resolution than {\it{Planck}}, which makes it possible to reveal the details of non-linear cross-correlations.

With such improvements, it is possible to use data from these future surveys to probe the full halo model for galaxies and the tSZ effect. The general halo model divides the power spectrum into the two-halo term, which accounts for the correlation between different halos, and the one-halo term, which accounts for correlations within the same halo, so that

\begin{equation}
    P_{UV}(k) = P_{UV, \mathrm{1h}}(k) + P_{UV, \mathrm{2h}}(k).
\end{equation}
Both terms are related to the profiles of $U$ and $V$ in Fourier space:

\begin{equation}
\begin{aligned}
P_{UV, \mathrm{1h}}(k) &=\int_0^{\infty} \dr M \frac{\dr n}{\dr M}\langle p_U(k \mid M) p_V(k \mid M)\rangle \\
P_{UV,\mathrm{2h}}(k) &=\langle b_U\rangle(k)\langle b_V\rangle(k) P^{\mathrm{lin}}(k) \\
\langle b_U\rangle(k) & \equiv \int_0^{\infty} \dr M \frac{\dr n}{\dr M} b_{\mathrm{h}}(M)\langle p_U(k \mid M)\rangle,
\end{aligned}
\end{equation}
where $P^{\mathrm{lin}}(k)$ is the linear power spectrum; $\dr n/\dr M$ is the halo mass function; $b_{\mathrm{h}}$ is the halo bias and $p_U(k \mid M)$ is the profile of the tracer $U$ with mass $M$ in Fourier space:
\begin{equation}
p_U(k \mid M) \equiv 4 \pi \int_{0}^{\infty} \dr r r^{2} \frac{\sin (k r)}{k r} p_U(r \mid M).
\end{equation}

For the one-halo term, one needs to calculate the correlation between different profiles. We take the one-parameter model from \citet{koukoufilippas2020tomographic} to account for the cross-correlation between abundances of $u$ and $v$:

\begin{equation}
    \left\langle p_U(k \mid M) p_V(k \mid M)\right\rangle=\left(1+\rho_{UV}\right)\left\langle p_U(k \mid M)\right\rangle\left\langle p_V(k \mid M)\right\rangle.
\end{equation}

The details of profiles have been introduced in Sect. \ref{sect:model}. For the HOD profile we still fix $\sigma_{\mathrm{M}}=0.15$ and $\alpha_\mathrm{s}=1$, and let $\{M_0, M_1, M_{\mathrm{min}}\}$ vary.

For the tSZ profile, we must note that the mass term in this formula is calibrated with X-ray observations and are possibly biased \citep{2014planckxx}. A `hydrostatic bias' $b_{\mathrm{H}}$ is introduced, so that the mass term in \eqref{eq:gnfw} is replaced by $(1-b_{\mathrm{H}})M$. In the modelling, we also multiply the power spectra with correction factor $R(k, z)$ defined in \eqref{eq:rkdef}. Now we can see that the non-linear bias $\cgk$ and $\cgy$ are directly related to $\rho_{\mathrm{GM}}$, $\rho_{\mathrm{GP}}$, and $b_{\mathrm{H}}$.

We construct the Fisher matrix:

\begin{equation}
    \mathrm{F}_{\alpha \beta} =\frac{\partial \boldsymbol{M}^{\mathrm{T}}}{\partial q_{\alpha} }\mathrm{Cov}^{-1}\frac{\partial \boldsymbol{M}}{\partial q_{\beta}},
\end{equation}
where $\boldsymbol{M}(q)$ is the cross-correlations given by full halo model described above. We also assume that the covariance matrix does not depend on parameters. The free parameters are $q\in \{ \log_{10} M_1, \log_{10} M_0, \log_{10} M_{\mathrm{min}}, b_{\mathrm{H}}, \rho_{\mathrm{GP}}, \rho_{\mathrm{GM}} \}$. The Fisher matrix is calculated at the best-fit parameter values, which are assumed to be $\{ 13, 11.68, 11.86, 0.16, -0.5, 0 \}$.

The hypothetical measurement should be extended to non-linear region, so we set the $\ell$ region to be $50<\ell<3000$. The Gaussian covariance matrix is calculated the same way as Eq. \eqref{eq:covg}, with $C^{uv}_{\ell}$ the real angular power spectra signal calculated from best-fit model plus noise spectrum:

\begin{equation}
    C^{uv}_{\ell} = C^{uv,\mathrm{signal}}_{\ell} + N^{uv}_{\ell}.
\end{equation}
For cross-correlations, $N^{uv}_{\ell}=0$. The galaxy count noise is taken as the shot noise:

\begin{equation}
    N^{\mathrm{gg}}_{\ell} = 1/\overline{N},
\end{equation}
where $\overline{N}$ is the mean number of galaxies per steradian on the survey. In this section, we present the forecast for the tomographic bin of a LSST/\euclid-like survey corresponding to the last tomographic bin of KiDS (0.9<$Z_\mathrm{B}\leq$1.2). The redshift distribution and galaxy numbers are from \citet{lsstsciencecollaboration2009lsst}. The information of LSST and \euclid \, surveys are summarised in Table.~\ref{table:lsst_euclid}. Since both has similar survey coverage and galaxy number density, we only show the forecast for LSST for clarity. We take the noise power spectra of CMB lensing and the tSZ effect for CMB-S4 presented in  \citet{Shirasaki_2019} and \citet{Schaan_2017}, respectively.

Since the hypothetical measurement extends into non-linear region, we include the non-Gaussian covariance matrix, which is given by:
\begin{equation}
\begin{aligned}
\operatorname{Cov}^{\mathrm{NG}}\left(C_{\ell}^{u v}, C_{\ell^{\prime}}^{w z}\right)&=\int_0^{\infty} d \chi \frac{W^{u}(\chi) W^{v}(\chi) W^{w}(\chi) W^{z}(\chi)}{4 \pi f_{\mathrm{sky}} \chi^{6}}\\
&\times T_{UVWZ}\left(k=\frac{\ell+1 / 2}{\chi}, k'=\frac{\ell'+1 / 2}{\chi}\right),
\end{aligned}
\end{equation}
where $T_{UVWZ}(k)$ is the trispectrum. Using the halo model, the trispectrum is decomposed into one- to four- halo terms. Here we only take the one-halo term into account since it dominates the scales we are interested in \citep{Pielorz_2010}:
\begin{equation}
\begin{aligned}
T_{UVWZ}^{\mathrm{1h}}(k, k') &\equiv \int_0^{\infty} \dr M \frac{\dr n}{\dr M} \\
&\times\langle p_U(k\mid M) p_V(k \mid M) p_W(k'\mid M) p_Z(k'\mid M)\rangle.
\end{aligned}
\end{equation}

\begin{figure*}
    \centering
    \includegraphics[width=0.8\textwidth]{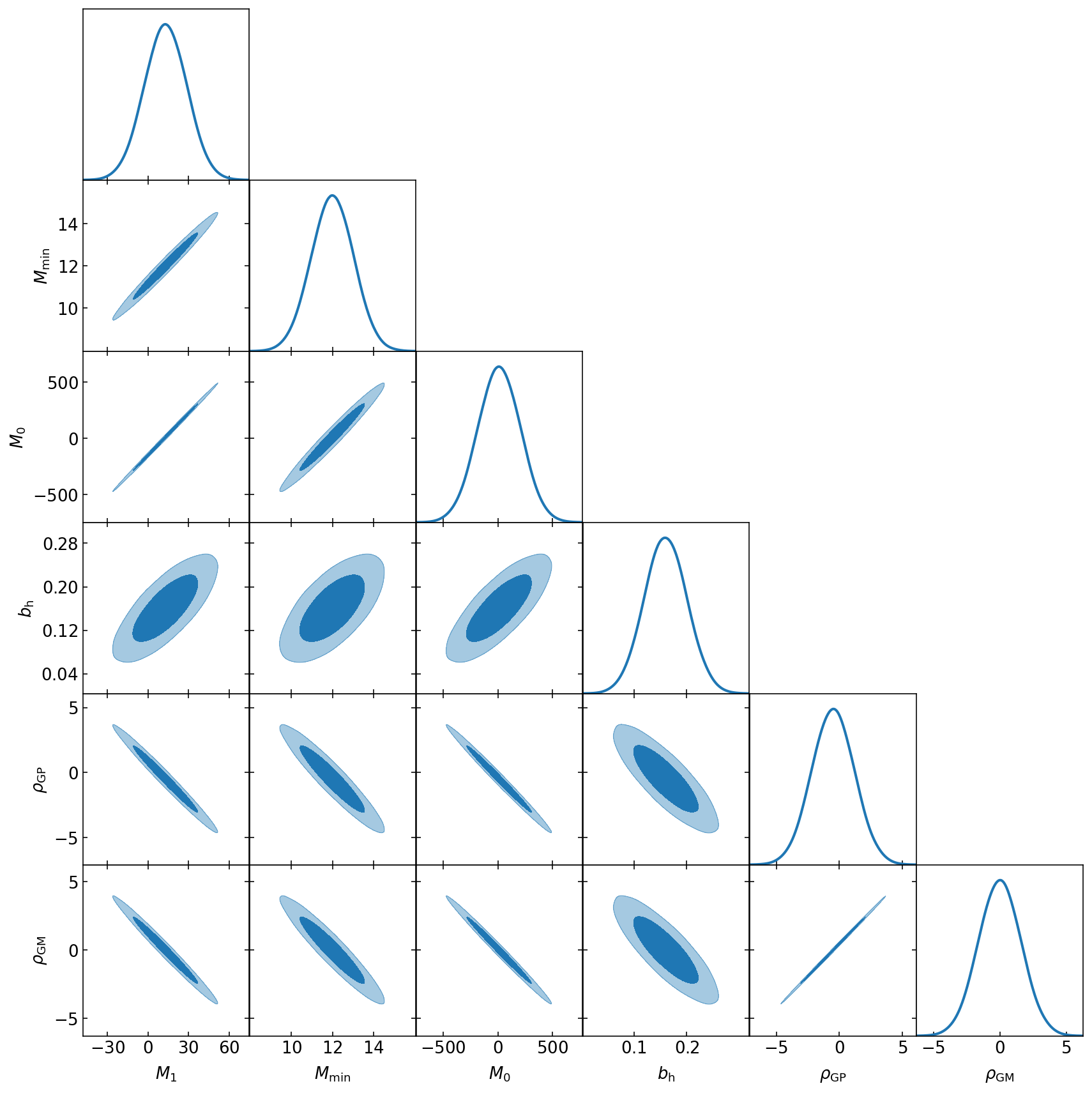}
    \caption{The 95\% and 68\% confidence contours of model parameters of the full halo model. The confidence contours are calculated from the Fisher matrix for LSST/\euclid -like galaxy surveys and CMB-S4.}
    \label{fig:forecast}
\end{figure*}

With these pieces of information, we calculate the covariance matrix as $\mathrm{Cov} = \mathrm{Cov^{\mathrm{G}}}+\mathrm{Cov^\mathrm{NG}}$. The confidence contours from the Fisher matrix is calculated and presented in Fig. \ref{fig:forecast}.

\begin{table}
\begin{tabular}{lllll}\toprule
Survey                 & Sky coverage {[}$\mathrm{deg}^2${]} & $f_{\mathrm{sky}}$ {[}\%{]} & $\overline{N}\, [\mathrm{arcmin^{-2}}]$ &  \\\midrule
LSST                   & $20000 $                             & 48.5                    & 55.5                                &  \\
\euclid & $15000$                             & 36.4                       & 37.0                                 & \\
\bottomrule
\end{tabular}
\caption{Information of LSST and \euclid $\,$ surveys needed for forecasting. We note that the sky coverage is the full sky coverage multiplied by the estimated overlapping fraction with CMB-S4.}
\label{table:lsst_euclid}
\end{table}

The parameter that we are interested in is $b_{\mathrm{H}}$, which is constrained as $0.16\pm 0.04$ for such LSST/\euclid-like surveys. The gas pressure bias $\bpe$ can then be calculated according to \eqref{eq:bpe_gnfw}, yielding a constraint $(0.37\pm 0.02)\,[\mathrm{meV/cm^3}]$. Although the constraining power for galaxy parameters $\{M_1, M_0, M_{\mathrm{min}}\}$ is weak, the constraining power for gas pressure bias is relatively strong, indicating that galaxy-CMB lensing cross-correlation is a valid method to break the degeneracy between parameters of galaxy distribution and other large-scale tracers when measuring galaxy cross-correlations. Our forecast indicates that future sky surveys like LSST and \euclid $\,$ as well as CMB-S4 will obtain data that can be used to conduct such measurements.

\end{appendix}

\end{document}